\documentclass[10pt,journal,twoside,final]{IEEEtran}

\usepackage{blindtext}
\usepackage{graphicx}
\usepackage{cite}
\usepackage{subfigure}
\usepackage{amssymb,amsmath}
\usepackage{amsthm}
\usepackage{mathtools}
\usepackage{multirow}
\usepackage{multicol}
\usepackage{verbatim}
\usepackage{color, soul}
\usepackage{enumitem}
\usepackage[normalem]{ulem}
\usepackage{lipsum}
\usepackage{cuted}
\usepackage{bm}
\usepackage{Maths}
\usepackage{adjustbox}
\usepackage{threeparttable}
\usepackage{booktabs}
\usepackage{array,tabularx,adjustbox,threeparttable, xparse}
\usepackage[table]{xcolor}
\usepackage[colorlinks=true,linkcolor=black,urlcolor=black,citecolor=black]{hyperref}
\newcolumntype{R}[1]{>{\raggedright\arraybackslash}p{#1}}
\usepackage[most]{tcolorbox}
\usepackage{tikz}
\usetikzlibrary{arrows.meta,positioning,fit}
\usepackage[switch,mathlines]{lineno}

\usepackage{dirtytalk}
\usepackage{algorithm}
\usepackage{algorithmic}

\makeatletter
\renewcommand{\@IEEEfigurecaptionsepspace}{\vskip -6pt\relax}
\makeatother

\newcommand{\C}{\mathbb{C}}

\DeclareRobustCommand{\basicfix}[1]{{\hypersetup{linkcolor=blue}\color{blue}#1}}

\begin{document}
\bstctlcite{IEEEfullAuthorNames}
	\title{ RIS Configuration with Low-Overhead Measurements via Physics-Guided Neural Networks}

	\author{
		Son~Dinh-Van, \IEEEmembership{Member,~IEEE},
		and Matthew~D. Higgins, \IEEEmembership{Senior Member,~IEEE}
		\IEEEcompsocitemizethanks{
			\IEEEcompsocthanksitem
			Son Dinh-Van and Matthew D. Higgins are with Warwick Manufacturing Group, University of Warwick, Coventry, CV4 7AL, U.K. Email: \{son.v.dinh, m.higgins\}@warwick.ac.uk.
			\IEEEcompsocthanksitem
			Corresponding author: Son~Dinh-Van (e-mail: son.v.dinh@warwick.ac.uk).
			\IEEEcompsocthanksitem
			This work was supported in part by the WMG Centre High Value Manufacturing Catapult, University of Warwick, Coventry, U.K.
		}
	}
	\maketitle
	\thispagestyle{empty}
	\pagestyle{empty}
	
	\begin{abstract}
		This paper studies reconfigurable intelligent surface (RIS) configuration for a single-input single-output (SISO) communication link using a two-stage estimate-then-optimize procedure. Within this framework, two common approaches to the estimation stage are pilot-based channel estimation and received signal strength (RSS)-based channel estimation. Our hardware experiments reveal substantial inter-block phase evolution consistent with residual CFO, while simulations show that such phase rotations can severely degrade pilot-based channel estimation when left uncompensated. In contrast, RSS-based approaches are less affected by these phase rotations, although they often require a large number of measurements to achieve a near-optimal received signal-to-noise ratio (SNR). To address this measurement requirement, we propose parametric physics-guided neural blind beamforming (\textsc{P-PGNBB}) for RSS-based channel estimation. Simulations and hardware experiments confirm that the proposed method achieves a received SNR close to that obtained with perfect channel state information (CSI) and comparable to existing methods, while requiring substantially fewer RSS measurements.
	\end{abstract}
	
	
	%-------------------------------------------------------------------------------------
	\section{Introduction}
	\label{se:intro}
	
	Reconfigurable intelligent surfaces (RISs) offer a promising means of improving wireless coverage by redirecting incident signals toward desired areas. This capability is realized by controlling the phase shifts introduced by the passive reflecting elements of the RIS \cite{basarAccess2019RIS,diRenzoJSAC2020smartRadio,liuCOMST2021RISSurvey,dinhvanJIOT2022}. However, the resulting performance gain depends critically on the selected reflection pattern. Therefore, determining a suitable RIS configuration is a central problem in RIS-assisted communications. Ideally, the configuration procedure should achieve a high received signal-to-noise ratio (SNR) with low acquisition and measurement overhead.
	
	One widely adopted approach to RIS configuration relies on \textit{pilot-based channel estimation}. Specifically, the transmitter sends known complex pilot sequences, while the receiver collects the corresponding observations under a sequence of RIS configurations. These observations are then used to estimate the channel and optimize the reflection states \cite{youJSAC2020channelEstimation}. This approach has attracted extensive research attention; see \cite{wuTCOM2021RISTutorial,zhengCOMST2022channelEstimationSurvey} and the references therein. Within this framework, existing methods improve acquisition through progressive channel refinement \cite{youJSAC2020channelEstimation} or exploit propagation geometry to reduce pilot overhead \cite{haghshenasTWC2024parametricCE}. Nevertheless, this approach faces three challenges in practice. First, pilot transmission, RIS switching, observation collection, and feedback require careful coordination and synchronization, which complicate integration into existing wireless networks. Moreover, the supporting integration procedures remain a subject of ongoing RIS standardization work \cite{etsiWorkProgramme2025_2026}. Second, the receiver must provide complex pilot observations, including their amplitude and phase, which may not be accessible through commodity hardware \cite{arunNSDI2020RFocus}. Third, in practice, carrier frequency offset (CFO) can rotate the observation associated with each RIS configuration by a different phase, even when the physical channel remains unchanged. Consequently, combining these observations without compensation can degrade channel estimation and the resulting configuration. Nevertheless, the effect of CFO is still not explicitly considered in some pilot-based channel-estimation studies \cite{youJSAC2020channelEstimation,haghshenasTWC2024parametricCE}. A rare example of work explicitly addressing CFO is the study in \cite{jeongTVT2023jointCFOCE}, which uses structured pilots and a correlation-based CFO estimator, followed by CFO compensation and channel estimation. However, this approach still requires suitable pilot design, processing of complex observations, and coordinated acquisition.
	
	An alternative approach, referred to as blind beamforming, configures the RIS using only received signal strength (RSS) feedback. By relying on scalar feedback, it reduces the need for access to complex receiver observations and can therefore simplify integration with existing hardware \cite{arunNSDI2020RFocus,dinhvanTWC2026BORN}. In addition, as discussed later in Section~\ref{se:phase_drift_channel_estimation}, RSS is less sensitive than complex pilot observations to the  phase rotations caused by CFO in the adopted model, thereby avoiding the need for phase alignment across probing configurations. Within this framework, existing methods can be divided into two families: \textit{model-free} and \textit{model-based} blind beamforming.
	
	Regarding the first family, \textit{model-free} blind beamforming selects reflection states directly from RSS measurements without learning an explicit channel model. For example, codebook search is a common approach because of its simplicity, selecting the strongest measured pattern from a predefined set \cite{peiTCOM2021RISPrototype}. In comparison, RFocus uses majority voting to determine binary states \cite{arunNSDI2020RFocus}, while conditional sample mean (CSM) selects each discrete phase state by comparing conditional received-power averages \cite{renTWC2023blindBeamforming}. Further approaches use element grouping \cite{laiTMC2025adaptiveBlind}, adaptive cross-entropy search \cite{chenWCL2024crossEntropy,binWCL2026adaptiveCrossEntropy}, or deterministic element-wise probing \cite{malekiArxiv2026limitedFeedback}. However, the achievable performance depends on the decision rules and, in some cases, the effectiveness of heuristic searches. In particular, the performance of codebook search is constrained by the codebook design, while methods such as RFocus and CSM can degrade when the background channel is weak \cite{laiTMC2025adaptiveBlind,dinhvanTWC2026BORN}. Thus, obtaining a configuration close to the optimal received SNR from a limited measurement budget remains challenging.
	
	In contrast, \textit{model-based} blind beamforming first estimates a channel-equivalent representation or an explicit RSS-response model and then optimizes the RIS configuration \cite{sunGLOBECOM2023PGNN,dinhvanTWC2026BORN,yanTWC2025powerAutocorrelation}. Hence, it also follows an estimate-then-optimize procedure, but utilizes RSS measurements instead of complex pilots. Consequently, it retains the acquisition advantages of RSS feedback, including the absence of an inter-configuration phase-reference requirement. For example, the single-layer neural network in \cite{sunGLOBECOM2023PGNN} represents received power through an element-wise channel model containing $2N+2$ trainable real parameters, where $N$ denotes the number of RIS elements. Hereafter, we refer to this method as \textsc{PGNN}. In our earlier work, \textsc{BORN} instead estimates a structured quadratic SNR model and provides a theoretical near-optimality guarantee for binary RIS configuration \cite{dinhvanTWC2026BORN}. This method demonstrates a significantly improved received SNR performance compared to model-free approaches. Alternatively, channel-autocorrelation estimation learns a power-response representation that supports multi-bit phase control \cite{yanTWC2025powerAutocorrelation}. Although the model-based techniques can achieve high received SNR, learning their representations from RSS measurements can require a substantial online measurement budget \cite{sunGLOBECOM2023PGNN,dinhvanTWC2026BORN,yanTWC2025powerAutocorrelation}.
	
	Despite these advances, each approach has practical limitations. Specifically, pilot-based channel estimation requires coordinated acquisition and synchronization, alongside with careful treatment of CFO, while model-free blind beamforming can leave a gap to the optimal received SNR. Meanwhile, existing model-based methods can require a large number of measurements to estimate a sufficiently accurate response model. To address these limitations, this paper develops a method that can attain near-optimal received SNR with low measurement overhead. By using RSS feedback, the method also avoids the need to estimate and compensate for phase rotations across probing configurations. Our method belongs to the model-based blind beamforming category.
	
	To this end, we consider a narrowband single-input single-output (SISO) link between a single-antenna transmitter (TX) and a single-antenna receiver (RX), assisted by an RIS with $N$ passive elements and $Q$ discrete phase states per element. Within this setting, we propose parametric physics-guided neural blind beamforming (\textsc{P-PGNBB}), which embeds an effective far-field channel representation into a shallow neural network. The key insight is that the cascaded channel coefficients across the RIS share a spatial structure determined by propagation. Accordingly, when the channel is adequately represented by $L$ effective components, these coefficients can be learned through the component gains and spatial frequencies instead of being treated as independent unknowns. Together with the background channel, the resulting model contains $4L+2$ trainable real parameters and therefore provides a compact representation when $L\ll N$. Using RSS measurements collected under known probing configurations, the network learns this representation online, after which the estimated channel is used to optimize the $Q$-ary reflection states. 
	%-------------------------------------------------------------------------------------
	\subsection{Contributions}
	
	The main contributions are summarized as follows. \textbf{Firstly}, we examine the phase-coherence requirement of pilot-based channel estimation through analysis, hardware measurements, and numerical simulations. In particular, the measurements reveal substantial phase evolution across successive pilot blocks, whereas the amplitude remains comparatively stable. These observations provide a practical motivation for an RSS-based approach without requiring phase alignment across RIS probing states. Our \textbf{second} contribution is the development of \textsc{P-PGNBB}, a shallow physics-guided neural network that learns a compact effective cascaded channel from RSS measurements using an $L$-component far-field representation. Specifically, we develop a deterministic initialization and an online training procedure, and combine the learned model with $Q$-ary configuration optimization. Its trainable parameter count is independent of $N$, thereby reducing the dimension of the model that must be learned from RSS measurements. \textbf{Finally}, we evaluate the received SNR and measurement requirements of \textsc{P-PGNBB} through simulations and hardware experiments. The simulations show that the method achieves average received SNR within $1.74$~dB of the perfect-CSI reference using $200$ RSS measurements. Furthermore, in the hardware experiment, it reaches a received SNR of $53$~dB using $400$ RSS measurements, compared with $1600$, $3200$, and $12800$ for \textsc{CSM}, \textsc{PGNN}, and \textsc{BORN}, respectively.
	
	\begin{figure*}[!t]
		\centering
		\includegraphics[width=\textwidth]{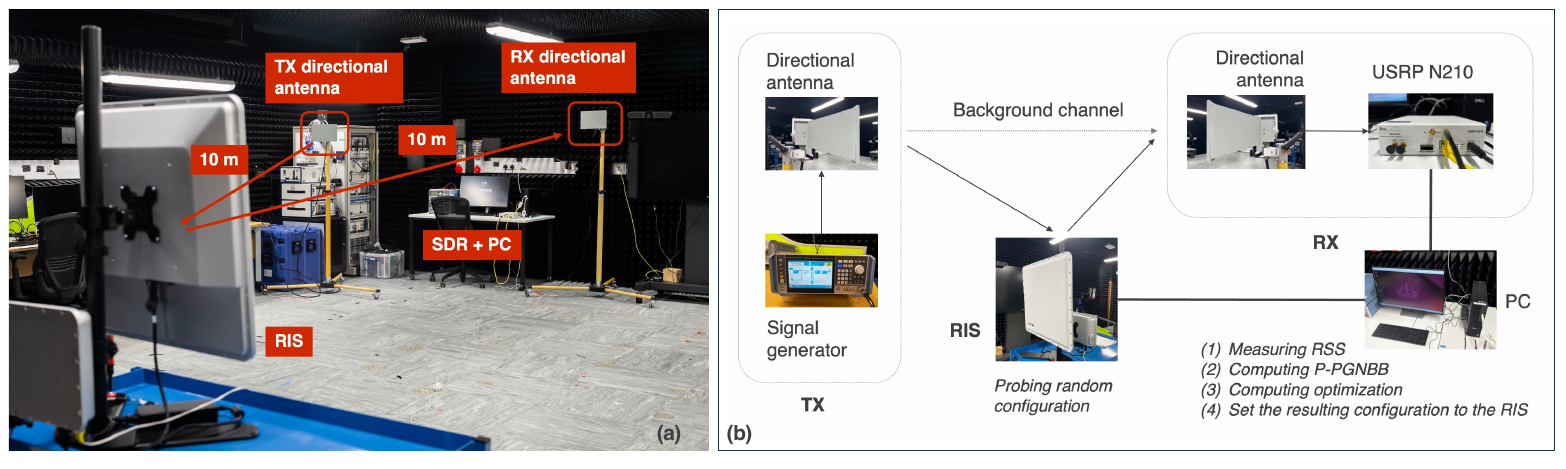}
		\caption{Experimental setup and RSS-based RIS configuration workflow: (a) laboratory testbed; (b) system model and processing steps for RSS acquisition, model training, and RIS configuration optimization.}
		\label{fig:experimental_setup}
	\end{figure*}
	
	\textit{Paper Organization:}
	Section~\ref{se:system_model_problem_formulation} introduces the system model and formulates the RIS configuration problem. Next, Section~\ref{se:phase_drift_channel_estimation} examines pilot-based channel estimation and its sensitivity to residual CFO. Building on this motivation, Section~\ref{se:parametric_pgnbb} develops the parametric channel representation and the proposed RSS-based learning method, while Section~\ref{se:q_ary_optimization} presents the optimization stage. Subsequently, Section~\ref{se:results} evaluates the method through simulations and hardware experiments. Finally, Section~\ref{se:conclusion} concludes the paper.
	
	\textit{Notations:}
	Scalars are denoted by regular letters, vectors by bold lowercase letters, and matrices by bold uppercase letters. A hat over a symbol denotes an estimate or prediction of the corresponding quantity. The symbols $(\cdot)^\top$, $(\cdot)^{\mathrm{H}}$, and $(\cdot)^+$ denote transpose, Hermitian transpose, and Moore--Penrose pseudoinverse, respectively. For $z\in\C$, $\Real{z}$, $\Imag{z}$, and $\abs{z}$ denote its real part, imaginary part, and magnitude. The imaginary unit is $\mathrm{j}\triangleq\sqrt{-1}$.
	%-------------------------------------------------------------------------------------
	
	\section{\texorpdfstring{System Model and Problem Formulation}{System Model and Problem Formulation}}
	\label{se:system_model_problem_formulation}
	
	\subsection{System Model}
	\label{subse:system_model}
	In this paper, we consider a narrowband communication link between a transmitter (TX) and receiver (RX), assisted by an RIS, as shown in Fig. \ref{fig:experimental_setup}(b). The RIS is equipped with $N$ passive elements with $Q\geq2$ discrete phase states per element. The end-to-end model is independent of a particular uplink or downlink protocol. Let $h_0\in\C$ denote the background channel, which includes all paths that do not involve the RIS, and $h_n\in\C$ denote the cascaded channel associated with the $n$-th RIS element.
	
	Let $\theta_n\in\{2\pi q/Q:q=0,\ldots,Q-1\}$ be the phase shift applied by the $n$-th RIS element. Define
	\begin{equation}
		\mathcal{X}_Q
		\triangleq
		\left\{e^{\mathrm{j}2\pi q/Q}:q=0,\ldots,Q-1\right\},
	\end{equation}
	and let $x_n\triangleq e^{\mathrm{j}\theta_n}\in\mathcal{X}_Q$ denote its unit-modulus reflection coefficient. The RIS configuration vector and cascaded channel vector are defined as
	\begin{equation}
		\bm{x}\triangleq[x_1,\ldots,x_N]^\top\in\mathcal{X}_Q^N,
		\qquad
		\bm{h}\triangleq[h_1,\ldots,h_N]^\top .
	\end{equation}
	Under an RIS configuration $\bm{x}$, the effective channel is $g(\bm{x})\triangleq h_0+\bm{x}^\top\bm{h}$. The received signal can be modeled as
	\begin{equation}
		Y(t)
	    =
		e^{\mathrm{j}\omega(t)}
		g(\bm{x})X(t)
		+
		Z(t),
	\end{equation}
	where $X(t)$ is the transmitted complex baseband signal and $Z(t)\sim\mathcal{CN}(0,\sigma^2)$ is the receiver noise. The term $\omega(t)$ is an unknown phase process caused by residual CFO, phase drift, or other synchronization imperfections. In the ideal coherent case, $\omega(t)=0$ across the training interval. This assumption is common in many RIS studies with pilot-based channel estimation. However, in Section~\ref{se:phase_drift_channel_estimation}, we will show why this assumption can be fragile in practice when pilot sequences are received sequentially over different RIS configurations.
	%-------------------------------------------------------------------------------------
	
	\vspace{-0.25cm}
	\subsection{Problem Formulation}
	\label{subse:problem_formulation}
	
	Our objective is to find the RIS state that maximizes the received SNR. Let $P\triangleq\E{\abs{X}^2}$ denote the transmit power, and define $\rho\triangleq P/\sigma^2$. Under RIS configuration $\bm{x}$, the received SNR is formulated as
	\begin{equation}
		\operatorname{SNR}(\bm{x})
		=
		\rho
		\abs{h_0+\bm{x}^\top\bm{h}}^2 .
	\end{equation}
	Our optimization problem can be formulated as
	\begin{equation}
		\mathrm{(P1)} \quad
		\underset{\bm{x}\in\mathcal{X}_Q^N}{\mathrm{maximize}} \quad
		\rho
		\abs{
		h_0+\bm{x}^\top\bm{h}
		}^{2}.
	\end{equation}
	Problem $\mathrm{(P1)}$ is fundamental in RIS-aided communications~\cite{wuTWC2019jointBeamforming,youJSAC2020channelEstimation,wuTCOM2021RISTutorial,zhengCOMST2022channelEstimationSurvey}. We follow the common estimate-then-optimize procedure, which has two stages:
	\begin{itemize}
		\item \textit{First stage:} This stage estimates the channel state information required by~$\mathrm{(P1)}$. We consider two approaches to this stage. The first one is \textit{pilot-based channel estimation}, which uses complex pilot observations to estimate $(h_0, \bm{h})$, as discussed in Section~\ref{se:phase_drift_channel_estimation}. The second approach is \textit{RSS-based channel estimation}, which utilizes RSS measurements to estimate an RSS-equivalent channel. Since RSS measurements discard the absolute phase, $h_0$ and $\bm{h}$ are identifiable only up to a common phase rotation. Nevertheless, this ambiguity does not affect RIS configuration because a common phase rotation leaves the objective in~$\mathrm{(P1)}$ unchanged. Related RSS-based methods similarly estimate equivalent channel or channel-autocorrelation representations~\cite{dinhvanTWC2026BORN,yanTWC2025powerAutocorrelation}. Section~\ref{se:parametric_pgnbb} develops the proposed RSS-based channel-estimation method.
		\item \textit{Second stage:} In the second stage, the RIS configuration is obtained by solving~$\mathrm{(P1)}$ using the estimated channel $(\widehat{h}_0, \widehat{\bm h})$, which will be presented in Section~\ref{se:q_ary_optimization}.
	\end{itemize}
	%-------------------------------------------------------------------------------------
	
	\section{Pilot-Based Channel Estimation and the Phase-Coherence Challenge}
	\label{se:phase_drift_channel_estimation}
	This section investigates pilot-based channel estimation from complex observations collected under different RIS configurations. Although numerous pilot-based channel-estimation methods have been developed for RIS-assisted systems, they commonly combine pilot observations associated with multiple RIS states, which are typically applied over successive pilot blocks. We therefore focus on a fundamental least-squares channel-estimation method as a representative example. Our objective is not to introduce a new channel estimator, but to use this example to highlight the phase-coherence requirement and the practical challenges caused by residual CFO.
	%-------------------------------------------------------------------------------------
	
	\subsection{Ideal Pilot-Based Channel Estimation}
	\label{subse:ideal_channel_estimation}
	
	Assume that there is no residual CFO or phase drift during the training measurements. For configuration $\bm{x}_k\in\mathcal{X}_Q^N$, the effective channel is $g_k
	= h_0+\bm{x}_k^\top\bm{h}$\basicfix{.} The TX sends a known pilot vector $\bm{s}_k\triangleq[X_{k,1},\ldots,X_{k,L_p}]^\top$ of length $L_p$. Under ideal phase coherence, the received pilot vector is written as
	\begin{equation}
		\bm{y}_k
		=
		\left(
		h_0+\bm{x}_k^\top\bm{h}
		\right)
		\bm{s}_k
		+
		\bm{z}_k,
	\end{equation}
	where $\bm{y}_k\triangleq[Y_{k,1},\ldots,Y_{k,L_p}]^\top$ and $\bm{z}_k\triangleq[Z_{k,1},\ldots,Z_{k,L_p}]^\top\sim\mathcal{CN}(\bm{0},\sigma^2\bm{I})$ is the noise vector. After pilot synchronization, the receiver correlates the received vector with the known pilot and obtains one complex scalar observation via
	\begin{equation}
		\label{eq:pilot_channel_observation}
		\widetilde{g}_k
		=
		\frac{\bm{s}_k^{\mathrm{H}}\bm{y}_k}{\bm{s}_k^{\mathrm{H}}\bm{s}_k}
		=
		h_0+\bm{x}_k^\top\bm{h}+e_k,
	\end{equation}
	where $e_k\triangleq\bm{s}_k^{\mathrm{H}}\bm{z}_k/(\bm{s}_k^{\mathrm{H}}\bm{s}_k)$ denotes the pilot-estimation error. Let $\widehat{h}_0$ and $\widehat{\bm{h}}$ denote the estimated background and cascaded channels, respectively. Define the augmented channel, its estimate, and the augmented probing vector as
	\begin{equation}
		\bm{q}\triangleq[h_0,\bm{h}^\top]^\top,
		\quad
		\widehat{\bm{q}}\triangleq
		[\widehat{h}_0,(\widehat{\bm{h}})^\top]^\top,
		\quad
		\bm{a}_k\triangleq[1,\bm{x}_k^\top]^\top.
	\end{equation}
	From \eqref{eq:pilot_channel_observation}, stacking $K$ RIS configurations yields
	\begin{equation}
	\begin{aligned}
		\widetilde{\bm{g}}
		&\triangleq
		[\widetilde{g}_1,\ldots,\widetilde{g}_K]^\top
		=
		\bm{A}\bm{q}+\bm{e},\\
		\bm{A}
		&\triangleq
		{
		\begin{bmatrix}
		\bm{a}_1 & \cdots & \bm{a}_K
		\end{bmatrix}^{\top}},
		\qquad
		\bm{e}\triangleq[e_1,\ldots,e_K]^\top .
	\end{aligned}
	\end{equation}
	The augmented channel contains $N+1$ complex unknowns. Therefore, at least $K\geq N+1$ probing configurations are required. Provided that $\bm{A}$ has full column rank, the least-squares channel estimate is
	\begin{equation}
		\widehat{\bm{q}}
		=
		\left(\bm{A}^{\mathrm{H}}\bm{A}\right)^{-1}
		\bm{A}^{\mathrm{H}}\widetilde{\bm{g}}
		=
		\bm{A}^{+}\widetilde{\bm{g}}.
	\end{equation}
	\textit{Optimization:} Based on these estimates, the RIS configuration is obtained by solving
	\begin{equation}
		\label{op: ideal_channel_estimation}
		\widehat{\bm{x}}
		\in
		\underset{\bm{x}\in\mathcal{X}_Q^N}{\arg\max}
		\abs{
			\widehat{h}_0+\bm{x}^\top\widehat{\bm{h}}
		}^2
		.
	\end{equation}
	With perfect channel estimates, the optimal RIS configuration can be obtained, as detailed in Section~\ref{se:q_ary_optimization}.
	%-------------------------------------------------------------------------------------
	
	\subsection{\texorpdfstring{Effect of CFO Across RIS States}{Effect of Phase Drift Across RIS Configurations}}
	\label{subse:phase_drift_model}
	
	In practice, the observation for RIS state $k$ may contain an unknown phase rotation in the hardware. Hence, the practical channel observation is
	\begin{equation}
		\widetilde{g}_k
		=
		e^{\mathrm{j}\omega_k}
		\left(
		h_0+\bm{x}_k^\top\bm{h}
		\right)
		+
		e_k.
	\end{equation}
	Stacking the observations gives the following equation:
	\begin{equation}
		\widetilde{\bm{g}}
		=
		\bm{D}_{\bm{\omega}}\bm{A}\bm{q}
		+
		\bm{e},
	\end{equation}
	where $\bm{\omega}\triangleq[\omega_1,\ldots,\omega_K]^\top$ and $\bm{D}_{\bm{\omega}}\triangleq\Diag{e^{\mathrm{j}\omega_1},\ldots,e^{\mathrm{j}\omega_K}}$. The least-squares estimator assumes
	$\widetilde{\bm{g}}\approx\bm{A}\bm{q}$, whereas the hardware observations follow
	$\widetilde{\bm{g}}\approx\bm{D}_{\bm{\omega}}\bm{A}\bm{q}$.
	If all measurements share a common phase rotation, the resulting global channel rotation does not affect the received-SNR maximization problem. In practice, however, the rotation may vary across measurements. This variation is harmful because it rotates the channel observation associated with each RIS configuration differently, thereby violating the phase coherence required for least-squares channel estimation. A common model for this phase rotation is~\cite{riceTCOM2001frequencyEstimation}
	\begin{equation}
	\omega_k
	\approx
	2\pi\delta_f t_k+\theta_0,
	\label{eq:cfo_phase_model}
	\end{equation}
	where $\delta_f$ denotes the residual CFO, $t_k$ is the observation time, and $\theta_0$ is the unknown initial phase offset. Without an external phase reference, $\theta_0$ is generally unidentifiable. Moreover, estimating $\delta_f$ requires additional observations and processing time. These requirements further complicate practical pilot-based channel estimation. However, we can see that this rotation disappears from RSS since
	\begin{equation}
		\label{eq:cfo_snr_invariance}
		\rho\abs{
			e^{\mathrm{j}\omega_k}
			\left(
			h_0+\bm{x}_k^\top\bm{h}
			\right)
		}^2
		=
		\rho\abs{
			h_0+\bm{x}_k^\top\bm{h}
		}^2.
	\end{equation}
	Therefore, in theory, the RSS associated with each RIS configuration is invariant to the CFO. In Section~\ref{se:results}, we experimentally examine this behavior in detail using the hardware testbed. This observation motivates RSS-based channel estimation, in which RSS measurements are used instead of complex-valued observations. However, it is very important to note that RSS measurements cannot identify the absolute channel phase. Nevertheless, an SNR-equivalent channel can still be estimated up to a common phase rotation. This ambiguity is immaterial for RIS configuration because a common rotation does not alter the underlying optimization problem.
	%-------------------------------------------------------------------------------------
	
	\section{RSS-Based Channel Estimation via P-PGNBB}
	\label{se:parametric_pgnbb}
	
	RSS-based channel estimation exploits the physical structure of the RSS to learn a channel representation without using pilot sequences.\footnote{Knowing the RSS is equivalent to knowing the received SNR when $\sigma^2$ is known. We therefore use the term ``RSS-based'' throughout this paper, even when the measurement labels are expressed in received SNR.} Existing model-based methods, including the physics-guided neural network (\textsc{PGNN})~\cite{sunGLOBECOM2023PGNN} and Blind Optimal RIS Beamforming with Sensing (\textsc{BORN})~\cite{dinhvanTWC2026BORN}, have demonstrated strong received-SNR performance but may require a substantial number of online RSS measurements. In this section, we propose parametric physics-guided neural blind beamforming (\textsc{P-PGNBB}), which estimates the channels by imposing a low-dimensional parametric channel structure. The estimated channel is identifiable only up to a common phase rotation, which is still sufficient for the optimization stage. Although this study focuses on far-field propagation, the proposed concept can be extended to near-field scenarios, which is left for future work.
	\vspace{-0.3cm}

	%-------------------------------------------------------------------------------------
	\subsection{Effective Far-Field Cascaded Channel Model}
	\label{subse:parametric_channel_model}

	Let the first RIS element be the spatial reference. For a planar RIS in the $yz$-plane, we write the known position of the $n$-th element as
	\begin{equation}
		\bm{p}_n
		\triangleq
		\begin{bmatrix}
			0 & i_n & k_n
		\end{bmatrix}^{\top},
		\qquad
		\bar{\bm{p}}_n
		\triangleq
		\begin{bmatrix}
			i_n & k_n
		\end{bmatrix}^{\top},
		\label{eq:ris_element_coordinates}
	\end{equation}
	with $\bm{p}_1=\bm{0}$ and $\bar{\bm{p}}_1=\bm{0}$. For a uniform planar RIS, $i_n$ and $k_n$ denote the horizontal and vertical offsets implied by the array spacings.
	%-------------------------------------------------------------------------------------
	
	\subsubsection{Single-component far-field model}

	First, consider one dominant far-field component on each side of the RIS. Let $\bm{p}_{\mathrm T}=r_{\mathrm T}\bm{u}_{\mathrm T}$ and $\bm{p}_{\mathrm R}=r_{\mathrm R}\bm{u}_{\mathrm R}$,
	where $r_{\mathrm T}\triangleq\Norm{\bm{p}_{\mathrm T}}_2$ and $r_{\mathrm R}\triangleq\Norm{\bm{p}_{\mathrm R}}_2$ are the distances from the reference element to the TX and RX, respectively. The vectors $\bm{u}_{\mathrm T}\triangleq\bm{p}_{\mathrm T}/r_{\mathrm T}$ and $\bm{u}_{\mathrm R}\triangleq\bm{p}_{\mathrm R}/r_{\mathrm R}$ are unit direction vectors that point from the reference element toward the TX and RX, respectively, and satisfy $\Norm{\bm{u}_{\mathrm T}}_2=\Norm{\bm{u}_{\mathrm R}}_2=1$. The exact distances through element $n$ are
	\begin{equation}
		r_{\mathrm T,n}
		=
		\Norm{\bm{p}_{\mathrm T}-\bm{p}_n}_2,
		\qquad
		r_{\mathrm R,n}
		=
		\Norm{\bm{p}_{\mathrm R}-\bm{p}_n}_2 .
	\end{equation}
	For the far-field channel model, we have that
	\begin{equation}
	\begin{aligned}
		r_{\mathrm T,n} &\approx r_{\mathrm T}-\bm{u}_{\mathrm T}^{\top}\bm{p}_n,\qquad r_{\mathrm R,n} &\approx r_{\mathrm R}-\bm{u}_{\mathrm R}^{\top}\bm{p}_n.
	\end{aligned}
	\label{eq:far_field_distance_approximation}
	\end{equation}
	Thus the excess path length relative to the reference element is written as
	\begin{equation}
		(r_{\mathrm T,n}+r_{\mathrm R,n})
		-
		(r_{\mathrm T}+r_{\mathrm R})
		\approx
		-
		(\bm{u}_{\mathrm T}+\bm{u}_{\mathrm R})^{\top}\bm{p}_n .
		\label{eq:far_field_excess_path}
	\end{equation}
	The cascaded channel coefficient through element $n$ is
	$h_n=\beta_n e^{-\mathrm{j}k_0(r_{\mathrm T,n}+r_{\mathrm R,n})}$, where $k_0\triangleq2\pi/\lambda$, $\lambda$ is the carrier wavelength, and $\beta_n$ is the channel gain. In the far field, $\beta_n\approx\beta$. Defining the reference-element coefficient $\gamma\triangleq\beta e^{-\mathrm{j}k_0(r_{\mathrm T}+r_{\mathrm R})}$ and using~\eqref{eq:far_field_excess_path} gives
	\begin{equation}
	\begin{aligned}
		h_n
		&\approx
		\gamma
		e^{\mathrm{j}k_0
		(\bm{u}_{\mathrm T}+\bm{u}_{\mathrm R})^\top
		\bm{p}_n}.
	\end{aligned}
	\label{eq:single_component_cascaded_product}
	\end{equation}
	Because the RIS lies in the $yz$-plane, only the two transverse components of $\bm{u}_{\mathrm T}+\bm{u}_{\mathrm R}$ affect the phase progression. Define
	\begin{equation}
	\begin{aligned}
		\kappa_y
		&\triangleq
		k_0(u_{\mathrm T,y}+u_{\mathrm R,y}), \ \kappa_z
		\triangleq
		k_0(u_{\mathrm T,z}+u_{\mathrm R,z}),\\
		\boldsymbol{\kappa}
		&\triangleq
		\begin{bmatrix}
			\kappa_y & \kappa_z
		\end{bmatrix}^{\top}.
	\end{aligned}
	\label{eq:single_component_parameter_definitions}
	\end{equation}
	The single-component cascaded channel is then
	\begin{equation}
		h_n
		=
		\gamma
		e^{\mathrm{j}\boldsymbol{\kappa}^{\top}\bar{\bm{p}}_n}.
		\label{eq:single_component_parametric_channel}
	\end{equation}
	Here, $\boldsymbol{\kappa}\in\R^2$ contains the horizontal and vertical spatial frequencies of the effective cascaded component. Define the steering vector $[\bm{b}(\boldsymbol{\kappa})]_n
	\triangleq
	e^{\mathrm{j}\boldsymbol{\kappa}^{\top}\bar{\bm{p}}_n}$, with
	$\bm{b}(\boldsymbol{\kappa})\in\C^N$\basicfix{.} The channel vector is therefore written as $\bm{h}=\gamma\bm{b}(\boldsymbol{\kappa})$. 
	%-------------------------------------------------------------------------------------
	
	\subsubsection{Multi-component far-field model}

	For multiple far-field components, let $L_{\mathrm T}$ and $L_{\mathrm R}$ denote the numbers of components on the TX--RIS and RIS--RX links, respectively. The cascaded coefficient through element $n$ is
	\begin{equation}
		\maxsizebox{\columnwidth}{!}{$\displaystyle
		\medmuskip=2mu \thickmuskip=3mu
		h_n =
		\biggl(\sum_{p=1}^{L_{\mathrm T}}
		\alpha_{\mathrm T,p}e^{\mathrm{j}k_0\bm{u}_{\mathrm T,p}^{\top}\bm{p}_n}\biggr) \times
		\!\biggl(\sum_{q=1}^{L_{\mathrm R}}
		\alpha_{\mathrm R,q}e^{\mathrm{j}k_0\bm{u}_{\mathrm R,q}^{\top}\bm{p}_n}\biggr)
		\approx \sum_{\ell=1}^{L}
		\gamma_\ell e^{\mathrm{j}\boldsymbol{\kappa}_\ell^{\top}\bar{\bm{p}}_n},
		$}
		\label{eq:multi_component_cascaded_product}
	\end{equation}
	where $\alpha_{\mathrm T,p}$ and $\alpha_{\mathrm R,q}$ are the corresponding link coefficients at the reference element. Each retained component $\ell$ represents a pairing of TX--RIS component $p$ and RIS--RX component $q$, with effective gain $\gamma_\ell$ and spatial frequency $\boldsymbol{\kappa}_\ell\triangleq k_0[u_{\mathrm T,p,y}+u_{\mathrm R,q,y},u_{\mathrm T,p,z}+u_{\mathrm R,q,z}]^\top$. Components with equivalent spatial responses are combined. Consequently, the cascaded channel is represented as
	\begin{equation}
		\bm{h}
		=
		\sum_{\ell=1}^{L}
		\gamma_\ell
		\bm{b}(\boldsymbol{\kappa}_\ell).
		\label{eq:multi_component_parametric_channel}
	\end{equation}
	For each component $\ell$, $\gamma_\ell\in\C$ is an unknown effective cascaded gain and $\boldsymbol{\kappa}_\ell=[\kappa_{y,\ell},\ \kappa_{z,\ell}]^{\top}\in\R^2$ is an unknown spatial-frequency vector.
	
	%-------------------------------------------------------------------------------------
	
	\subsubsection{Parametric SNR Model}

	Substituting the parametric channel into the SNR model yields
	\begin{equation}
		\operatorname{SNR}(\bm{x})
		=
		\rho
		\abs{
		h_0+
		\sum_{\ell=1}^{L}
		\gamma_\ell
		\bm{x}^\top\bm{b}(\boldsymbol{\kappa}_\ell)
		}^2.
		\label{eq:parametric_snr_model}
	\end{equation}
	
	\begin{remark}(\textbf{The number of unknown parameters})
		\label{rm:num_parameters}
		As can be seen, each path is parameterized by $\gamma_l\in\mathbb{C}$ and $\boldsymbol{\kappa}_l\in\mathbb{R}^2$. Therefore, together with $h_0$, the model contains $4L+2$ unknown real parameters. In the special case of a single dominant path, i.e., $L=1$, only $6$ real-valued unknown parameters need to be estimated.
	\end{remark}
	%-------------------------------------------------------------------------------------
	
	\subsection{Parametric Physics-Guided Neural Blind Beamforming}
	\label{subse:parametric_neural_rss_model}

	As discussed in Remark~\ref{rm:num_parameters}, a key advantage of the parametric channel model is its reduced number of unknown parameters. Based on~\eqref{eq:parametric_snr_model}, the mapping from the RIS configuration $\bm{x}$ to the received SNR can be represented by a physics-guided neural network with only $4L+2$ trainable parameters. The resulting network is shallow rather than deep and can therefore be trained efficiently using a relatively small number of RSS measurements. We refer to this network as parametric physics-guided neural blind beamforming (\textsc{P-PGNBB}). Its trainable parameter set is given by
	\begin{equation}
	\Theta
	\triangleq
	\left\{
	h_0,
	\left\{
	\gamma_{\ell},
	\boldsymbol{\kappa}_{\ell}
	\right\}_{\ell=1}^{L}
	\right\}.
	\end{equation}
	
\begin{figure}[!t]
		\centering
		\resizebox{\columnwidth}{!}{%
		\begin{tikzpicture}[
			font=\sffamily\scriptsize,
			>=Latex,
			known/.style={draw, rounded corners=1.5mm, fill=gray!15, align=center, minimum height=9mm, inner sep=1.5mm},
			train/.style={draw=blue!55!black, rounded corners=1.5mm, fill=blue!10, align=center, minimum height=9mm, inner sep=1.5mm},
			physics/.style={draw, rounded corners=1.5mm, fill=white, align=center, minimum height=9mm, inner sep=1.5mm},
			inner/.style={draw=gray!55, rounded corners=1mm, fill=gray!5, align=center, minimum height=10mm, inner sep=1mm},
			output/.style={draw=teal!60!black, rounded corners=1.5mm, fill=teal!8, align=center, minimum height=9mm, inner sep=1.5mm},
			loss/.style={draw, rounded corners=1.5mm, fill=gray!8, align=center, minimum height=9mm, inner sep=1.5mm},
			band/.style={font=\sffamily\tiny, text=gray!70!black, anchor=west},
			flow/.style={->, line width=0.5pt},
			update/.style={->, densely dotted, draw=black, line width=0.8pt}
		]
			\node[band] at (-4.1,5.35) {MODEL INPUTS};
			\node[known, text width=25mm] (input) at (-2.3,4.65) {RIS configuration\\[0.5mm]$\bm{x}\in\mathcal{X}_Q^N$};
			\node[train, text width=40mm] (parameters) at (1.55,4.65) {Trainable parameters\\[0.5mm]$\Theta=\{h_0,\gamma_\ell,\bm\kappa_\ell\}_{\ell=1}^{L}$};

			\node[physics, minimum width=82mm, minimum height=20mm] (branches) at (0,2.75) {};
			\node[band] at (-3.95,3.55) {SPATIAL COMPONENTS, $\ell=1,\ldots,L$};
			\node[inner, text width=20mm] (steering) at (-2.7,2.7) {Steering vector\\$\bm b_\ell=\bm b(\bm\kappa_\ell)$};
			\node[inner, text width=20mm] (projection) at (0,2.7) {RIS projection\\$z_\ell=\bm{x}^{\top}\bm b_\ell$};
			\node[inner, text width=20mm] (component) at (2.7,2.7) {Component field\\$q_\ell=\gamma_\ell z_\ell$};
			\node[physics, text width=28mm] (sum) at (-1.65,0.75) {Coherent sum\\$g=h_0+\sum_{\ell=1}^{L}q_\ell$};
			\node[physics, text width=22mm] (magnitude) at (-1.65,-0.65) {SNR scaling\\$\rho\abs{g}^{2}$};
			\node[output, text width=24mm] (prediction) at (1.65,-0.65) {Predicted RSS\\$\widehat r(\bm{x};\Theta)$};

			\node[band] at (-4.1,-1.55) {ONLINE TRAINING};
			\node[known, text width=24mm] (measurement) at (-1.65,-2.25) {Measured RSS\\$r$};
			\node[loss, text width=27mm] (mse) at (1.65,-2.25) {Prediction loss\\$\mathcal{L}(\Theta)=\operatorname{MSE}(\widehat r,r)$};

			\draw[flow] (input.south) -- (input.south |- branches.north);
			\draw[flow] (parameters.south) -- (parameters.south |- branches.north);
			\draw[flow] (steering) -- (projection);
			\draw[flow] (projection) -- (component);
			\draw[flow] ([xshift=-1.65cm]branches.south) -- (sum.north);
			\draw[flow] (parameters.east) -- ++(6mm,0) |- node[pos=0.8, above] {$h_0$} (sum.east);
			\draw[flow] (sum.south) -- (magnitude.north);
			\draw[flow] (magnitude) -- (prediction);
			\draw[flow] (prediction.south) -- (mse.north);
			\draw[flow] (measurement) -- (mse);
			\draw[update] (mse.east) -- ++(18mm,0) |- ([yshift=4mm]parameters.north) -| (parameters.north);
			\node[font=\sffamily\scriptsize\itshape, text=black, rotate=90] at (5.25,1.15) {Parameter updates via backpropagation};
		\end{tikzpicture}
		}
		\caption{The architecture of \textsc{P-PGNBB}.}
		\label{fig:parametric_pgnbb_architecture}
	\end{figure}
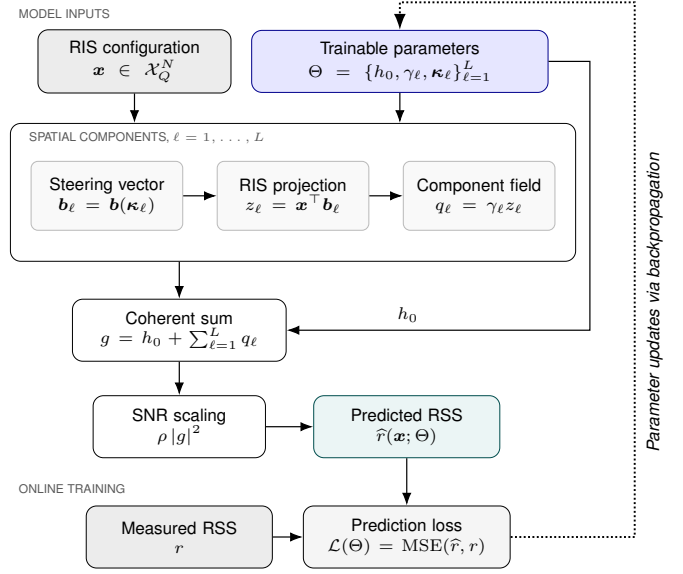

	The proposed \textsc{P-PGNBB} architecture is illustrated in Fig.~\ref{fig:parametric_pgnbb_architecture}. It can be implemented using standard machine learning libraries, such as $\mathtt{PyTorch}$ or $\mathtt{TensorFlow}$.

	%-------------------------------------------------------------------------------------
	
	\subsection{Online Training and Initialization}
	\label{subse:parametric_online_training}
	\subsubsection{Online Training}
	\label{subsubse:parametric_online_training}

	During each online configuration interval, the RIS applies $K$ random probing configurations, following a procedure similar to those in~\cite{renTWC2023blindBeamforming,sunGLOBECOM2023PGNN,dinhvanTWC2026BORN}, and collects the dataset $\mathcal{D}
	\triangleq
	\left\{
	(\bm{x}_k,r_k)
	\right\}_{k=1}^{K},$
	where the entries of each $\bm{x}_k$ are drawn independently and uniformly from $\mathcal{X}_Q$, and $r_k$ denotes the corresponding received SNR. Hence, each training sample consists solely of an RIS configuration and its measured RSS, without requiring CSI.
	
	The proposed \textsc{P-PGNBB} model is trained online by minimizing the following mean squared error (MSE) loss:
	\begin{equation}
	\begin{aligned}
	\mathcal{L}_{\mathrm{MSE}}(\Theta;\mathcal{D})
	&\triangleq
	\frac{1}{K}
	\sum_{k=1}^{K}
	\left[
	r_k
	-
	\widehat r(\bm{x}_k;\Theta)
	\right]^2,\\
	\widehat{\Theta}
	&\in
	\underset{\Theta}{\argmin}\;
	\mathcal{L}_{\mathrm{MSE}}(\Theta;\mathcal{D}).
	\end{aligned}
	\label{eq:parametric_training_loss}
	\end{equation}
	Since the loss function is differentiable with respect to $\Theta$, its gradients can be computed through backpropagation and the model can be trained using standard gradient-based optimizers, such as $\mathtt{Adam}$ or $\mathtt{L\mbox{-}BFGS}$. It is worth noting that training is performed online at deployment, after the measurement dataset has been collected, rather than via offline pretraining. This is computationally feasible because $\Theta$ is a small set, as mentioned in Remark \ref{rm:num_parameters}.
	%-------------------------------------------------------------------------------------
	
	\subsubsection{Initialization}
	\label{subsubse:parametric_initialization}
	The loss function shown in~\eqref{eq:parametric_training_loss} is typically nonconvex; hence, its solution can be sensitive to the starting point. We therefore use the deterministic initialization, as shown in Algorithm~\ref{alg:ppgnbb_initialization}. The detailed explanation of the initialization approach is provided in Appendix~\ref{app:ppgnbb_initialization}. The returned parameter set $\Theta^{(0)}$ is used as the starting point for training by solving~\eqref{eq:parametric_training_loss}. Both initialization and training use only the RSS measurements. Once \textsc{P-PGNBB} has been trained, its model parameters are estimated, and an SNR-equivalent cascaded channel can be constructed as $\widehat{\bm h} \triangleq \sum_{\ell=1}^{L} \widehat{\gamma}_{\ell} \bm b\!\left(\widehat{\boldsymbol{\kappa}}_{\ell}\right).$
	We emphasize again that the pair $(\widehat h_0,\widehat{\bm h})$ is identifiable only up to a common phase rotation. This ambiguity does not affect the objective function $\mathrm{(P1)}$.

	\begin{algorithm}[t]
		\caption{\textsc{P-PGNBB Initialization}}
		\label{alg:ppgnbb_initialization}
		\begin{algorithmic}[1]
			\item[\textbf{Input:}] $\mathcal D$, $L, \mathcal G_y,\mathcal G_z$.
			\item[\textbf{Output:}] $\Theta^{(0)}=\{h_0^{(0)},\{\gamma_\ell^{(0)},\bm\kappa_\ell^{(0)}\}_{\ell=1}^{L}\}$
			\STATE Set $h_0^{(0)}\gets0$ and form $\mathcal G\gets\mathcal G_y\times\mathcal G_z$.
			\STATE Apply~\eqref{eq:grid_candidate_snr_response}--\eqref{eq:grid_frequency_initialization} to $\mathcal D$ over $\mathcal G$ to obtain $\bm\kappa_1^{(0)}$ and $\gamma_1^{(0)}$.
			\IF{$L=1$}
				\STATE \textbf{return} $\Theta^{(0)}$.
			\ENDIF
			\STATE Set the first provisional gain $\widetilde\gamma_1^{(0)}\gets\gamma_1^{(0)}$.
			\FOR{$\ell=2$ to $L$}
				\STATE Evaluate $\widehat r_k^{(\ell-1)}$ for all $k$ using~\eqref{eq:multicomponent_initial_prediction}.
				\STATE Form $\mathcal D_{\mathrm{res}}^{(\ell)}\gets\{(\bm x_k,r_{k,\mathrm{res}}^{(\ell)})\}_{k=1}^{K}$ using~\eqref{eq:multicomponent_initial_residual}.
				\STATE Apply~\eqref{eq:grid_candidate_snr_response}--\eqref{eq:grid_frequency_initialization} to $\mathcal D_{\mathrm{res}}^{(\ell)}$ and obtain $\bm\kappa_\ell^{(0)}$ and $\widetilde\gamma_\ell^{(0)}$.
			\ENDFOR
			\STATE Form~\eqref{eq:multicomponent_lifted_gain_matrix}--\eqref{eq:multicomponent_lifted_prediction} and solve~\eqref{eq:multicomponent_hermitian_fit} using $\mathcal D$ to obtain $\widetilde{\bm G}$.
			\STATE Extract $\bm\gamma^{(0)}$ using~\eqref{eq:multicomponent_gain_initialization}.
			\STATE \textbf{return} $\Theta^{(0)}$.
		\end{algorithmic}
	\end{algorithm}

	\begin{remark}[\textbf{Computational complexity of Algorithm~\ref{alg:ppgnbb_initialization}}]
		\label{rm:initialization_complexity}
		Let $M\triangleq\abs{\mathcal G}=\abs{\mathcal G_y}\abs{\mathcal G_z}$ denote the grid size. For $L=1$, the total cost for initialization is $O(KNM)$. For $L>1$, the total cost is $O(KNM+LKM+KL^4+L^6)$. It is linear in $K$, $N$, and $M$. Note that $L$ is predefined and typically small in many scenarios. Later, we will show that a small $L$ can still achieve good SNR in Section \ref{se:results}.
	\end{remark}
	
	%-------------------------------------------------------------------------------------
	\section{\texorpdfstring{Optimization for \(Q\)-Ary RIS}{Optimization for Q-Ary RIS}}
	\label{se:q_ary_optimization}
	We now turn to \textit{Stage 2}, which optimizes the RIS configuration using the estimated channel. The same procedure applies to both pilot-based and RSS-based channel estimation. Specifically, we seek to solve
	\begin{equation}
		\mathrm{(P2)} \quad
		\underset{\bm{x}\in\mathcal X_Q^N}{\mathrm{maximize}}
		\quad
		\abs{
			\widehat h_0
			\mathbin{+}
			\bm{x}^{\top}\widehat{\bm h}
		}^{2}.
		\label{eq:parametric_configuration_objective}
	\end{equation}

	Let $\widehat{\bm{x}}$ denote a solution of~$\mathrm{(P2)}$. Although its feasible set contains $Q^N$ configurations, an optimal solution can be found efficiently using a directional event sweep inspired by~\cite{pekcanOJCOMS2024elementwiseRIS}. For a fixed direction $\phi$, each RIS element independently selects the phase state whose contribution has the largest projection onto that direction. This selection changes only when $\phi$ crosses one of at most $NQ$ phase boundaries. Algorithm~\ref{alg:qary_event_sweep} summarizes the main steps while the detailed explanation is provided in Appendix~\ref{app:qary_optimization_derivation}.

	\begin{algorithm}[!t]
		\caption{\textsc{Q-ary optimization}}
		\label{alg:qary_event_sweep}
		\begin{algorithmic}[1]
			\item[\textbf{Input:}] $\widehat h_0,\widehat{\bm h}$.
			\item[\textbf{Output:}] $\widehat{\bm{x}}$.
			\STATE Compute the boundaries via~\eqref{eq:qary_phase_boundaries}. Then sort them by angle and group coincident boundaries as $\mathcal E_1,\ldots,\mathcal E_B$.
			\STATE Choose a starting angle $\phi_0$ in the interval that goes from the largest boundary to the smallest boundary. Then initialize $\bm{x}$ using~\eqref{eq:qary_projection_selection}.
			\STATE Compute $\widehat g\gets\widehat h_0+\bm{x}^{\top}\widehat{\bm h}$ and set $(\widehat{\bm{x}},p^{\star})\gets(\bm{x},\abs{\widehat g}^2)$ according to~\eqref{eq:parametric_configuration_objective}.
			\FOR{$b=1$ to $B$}
				\STATE For each element $n$ associated with $\mathcal E_b$, update $x_n$ using~\eqref{eq:qary_projection_selection}. Leave all other elements unchanged.
				\STATE Update $\widehat g$ incrementally and evaluate $p\gets\abs{\widehat g}^2$ using~\eqref{eq:parametric_configuration_objective}. If $p>p^{\star}$, set $(\widehat{\bm{x}},p^{\star})\gets(\bm{x},p)$.
			\ENDFOR
			\STATE \textbf{return} $\widehat{\bm{x}}$.
		\end{algorithmic}
	\end{algorithm}

	\begin{remark}[\textbf{Optimality and computational complexity of Algorithm~\ref{alg:qary_event_sweep}}]
		As shown in Appendix~\ref{app:qary_optimization_derivation}, Algorithm~\ref{alg:qary_event_sweep} returns a globally optimal solution of~$\mathrm{(P2)}$. Its overall time complexity is $O\!\left(NQ\log(NQ)\right)$, dominated by sorting at most $NQ$ boundaries.
	\end{remark}
	
	%-------------------------------------------------------------------------------------
	
	\section{\texorpdfstring{Experimental and Simulation Results}{Experimental and Simulation Results}}
	\label{se:results}
	In this section, we first characterize the effect of CFO through hardware measurements. We then benchmark the proposed \textsc{P-PGNBB} against existing methods using both hardware experiments and simulations. The primary objective is to assess whether \textsc{P-PGNBB} can approach the upper-bound performance with fewer online measurements than existing algorithms require.
	
	In addition to \textsc{P-PGNBB}, we consider 8 methods:\footnote{Given the extensive literature on pilot-based channel estimation and blind beamforming, an exhaustive comparison is beyond the scope of this study. We select representative benchmarks, including commonly used approaches such as codebook search and CSM. BORN is also selected owing to its theoretical near-optimality guarantee. We choose PGNN and pilot-based channel estimation for their close relevance to our approach. Moreover, Perfect CSI provides a performance upper bound in simulations. These benchmarks offer insights into the tradeoff between measurement overhead and received-SNR performance.}
	\begin{itemize}
		\item \textit{Perfect CSI:} RIS configuration optimized using perfect CSI, serving as an upper bound on the received SNR in simulations.
		
		\item \textit{CE-Ideal:} RIS configuration obtained through pilot-based channel estimation in the absence of CFO.
		
		\item \textit{CE-Practical:} RIS configuration obtained through pilot-based channel estimation in the presence of CFO. In this case, the CFO is not compensated.
		
		\item \textsc{PGNN:} An RSS-based channel estimation method using a physics-guided neural network~\cite{sunGLOBECOM2023PGNN}.
		
		\item \textsc{BORN:} A model-based blind beamforming method using a quadratic model and channel-correlation estimation~\cite{dinhvanTWC2026BORN}.
		
		\item \textsc{CSM:} A model-free blind beamforming method using conditional sample means~\cite{renTWC2023blindBeamforming}.
		
		\item \textit{DFT Codebook:} RIS configuration obtained by searching discrete Fourier transform (DFT) codebooks containing 32 or 64 codewords~\cite{peiTCOM2021RISPrototype}.
		
		\item \textit{Zero-State:} All RIS phase shifts are set to zero.
	\end{itemize}
	
	For \textsc{P-PGNBB}, we use a $20 \times 20$ initialization grid. Both \textsc{P-PGNBB} and \textsc{PGNN} are trained using the $\mathtt{L\mbox{-}BFGS}$ optimizer, with learning rate \(\eta=10^{-2}\) and a maximum of \(T_{\max}=1000\) iterations. The number of measurements used by each method is specified for each simulation.
	%-------------------------------------------------------------------------------------
	\subsection{Experimental Setup}
	\label{subse:experimental_setup}
	 	We conduct the experiments in the Communications and Navigations Laboratory in the National Automotive Innovation Centre at the University of Warwick, U.K. As illustrated in Fig.~\ref{fig:experimental_setup}(a), the laboratory contains chairs, desks, and a wide range of equipment made from different materials, which can contribute to scattering. Within this setup, a Rohde \& Schwarz SMBV100B signal generator connected to a directional antenna at the TX transmits waveforms at a power of $-20$ dBm. Meanwhile, at the RX, an Ettus USRP N210 software-defined radio (SDR) receives the signals through a directional antenna connected to its second radio frequency (RF) port. The link is assisted by an RIS consisting of a $16\times16$ array of passive reflecting elements~\cite{tmytekWeb2025xrifle}, each of which has only two states. To coordinate RIS control, signal acquisition, and measurement processing, both the RIS and SDR are connected to a computer. Throughout the measurement campaign, the setup operates at a carrier frequency of $f_c=4.7$ GHz, while the TX, RX, and RIS remain at fixed positions, with both the TX--RIS and RIS--RX distances set to $10$ m. The SDR operates at a sampling rate of $5$ MHz, and the average noise floor at the RX is recorded at approximately $-104.48$ dBm.
	%-------------------------------------------------------------------------------------
	\subsection{Experimental Results}

	\subsubsection{Phase coherence measurements}
	\label{subse:results_phase_drift}
	
	\begin{figure}[!t]
		\includegraphics[width=\linewidth]{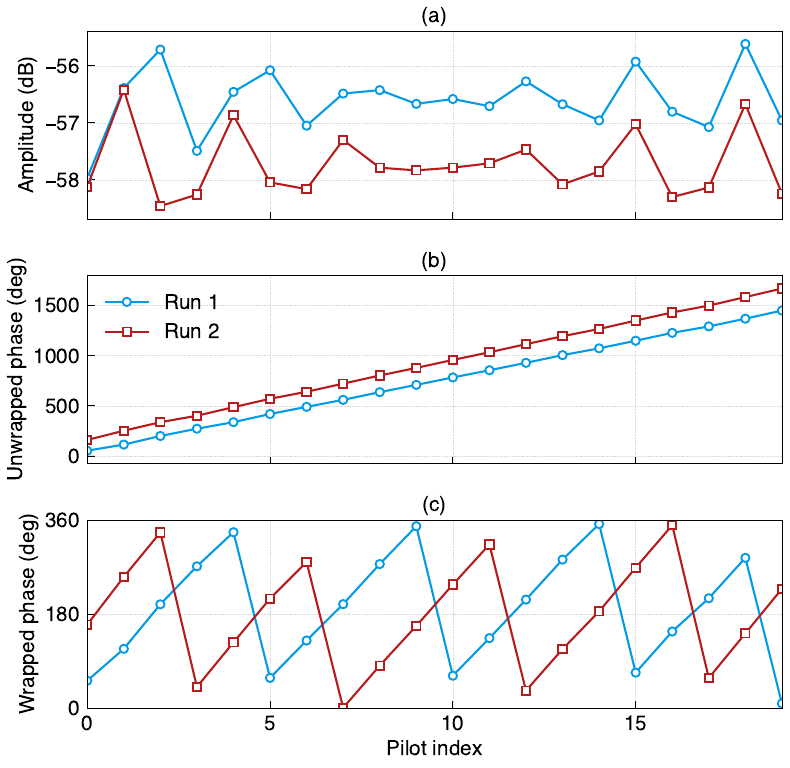}
		\caption{The complex channel estimated in the experiments using a wired connection: (a) amplitude, (b) unwrapped phase, and (c) wrapped phase.}
		\label{fig:cable_phase_drift}
	\end{figure}
	
	To characterize the CFO induced by the measurement hardware, the signal generator is connected directly to the SDR through a coaxial cable. This wired connection is used to remove fluctuations associated with the wireless propagation channel, allowing the phase evolution caused by the hardware to be isolated. The output power of the SMBV100B is reduced to $-50$ dBm to protect the SDR input. For this experiment, the signal generator repeatedly transmits a Zadoff--Chu (ZC) pilot sequence of length $L_p=1023$. At the RX, the direct current (DC) component is first removed from the captured in-phase and quadrature (I/Q) samples. The received pilot is then located by cross-correlating the samples with the known ZC sequence, after which the corresponding complex pilot observation is extracted for the CFO analysis.

	Fig.~\ref{fig:cable_phase_drift} shows the estimated amplitude and phase across successive pilot blocks for two independent measurement runs. As illustrated in Fig.~\ref{fig:cable_phase_drift}(a), the estimated amplitude remains comparatively stable throughout both runs, with standard deviations of approximately $0.55$ and $0.57$~dB for Runs~1 and~2, respectively. In contrast, the unwrapped phase in Fig.~\ref{fig:cable_phase_drift}(b) exhibits an approximately linear evolution, consistent with the CFO phase model shown in \eqref{eq:cfo_phase_model}. As we can see, from the $1$st to the $20$th pilot block, the accumulated phase changes are approximately $1396^\circ$ and $1509^\circ$, while linear fits yield average phase rotations of $73.4^\circ$ and $78.6^\circ$ per block for Runs~1 and~2, respectively. Consequently, the wrapped phase in Fig.~\ref{fig:cable_phase_drift}(c) exhibits periodic sawtooth discontinuities, despite the physical cable channel remaining nearly unchanged. Moreover, the two runs begin with different initial phases, indicating that the measurement chain preserves neither phase coherence across sequential pilot blocks nor a reproducible initial phase in the absence of an external reference.
	
	This experiment highlights two important insights for RIS configuration in practice. First, CFO must be considered in pilot-based channel estimation because it typically combines complex observations collected over multiple sequential pilot blocks. Second, the relative stability of the RSS measurements suggests that RSS-based channel estimation is considerably less sensitive to this effect, since it does not require phase coherence across pilot blocks and the CFO-induced phase rotation can be neglected when forming RSS measurements.
	%-------------------------------------------------------------------------------------
	\subsubsection{Benchmarks of \textsc{P-PGNBB}}
	\label{subse:results_rss_configuration}

	\begin{figure}[t]
		\centering
		\includegraphics[width=\linewidth]{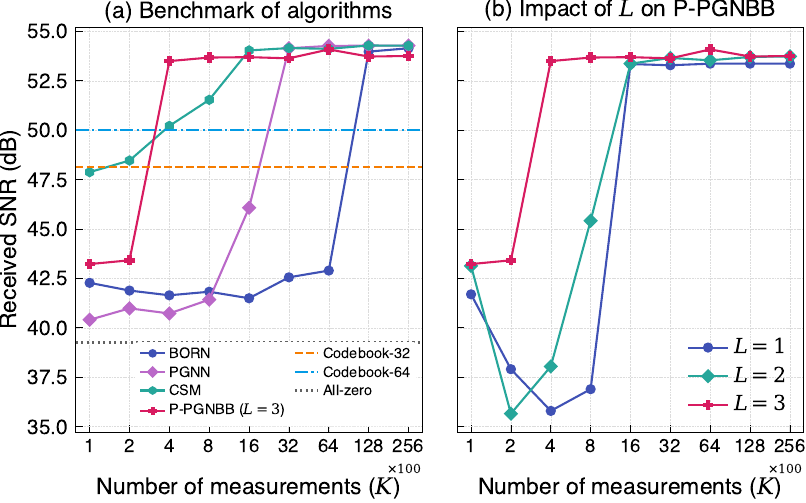}
		\caption{Received SNR versus the number of RSS measurements $K$ in the experiment: (a) Benchmarks of \textsc{P-PGNBB} ($L = 3$) against other algorithms; and (b) benchmark of \textsc{P-PGNBB} with $L\in\{1,2,3\}$.}
		\label{fig:hardware_alg_benchmark}
	\end{figure}

	We next evaluate the received SNR obtained with different numbers of RSS measurements. Fig.~\ref{fig:hardware_alg_benchmark}(a) benchmarks \textsc{P-PGNBB} with $L=3$ against the existing methods. As can be seen, the all-zero, codebook-32 and codebook-64 approaches achieve SNRs of $39.26$, $48.15$, and $50.00$~dB, respectively. Hence, using a codebook improves the received SNR significantly compared to the all-zero setting. By comparison, at $K=400$, \textsc{P-PGNBB} achieves $53.50$~dB, compared with $40.73$, $41.65$, and $50.22$~dB for \textsc{PGNN}, \textsc{BORN}, and \textsc{CSM}, respectively. Its gains over the all-zero configuration and the two codebook references are $14.24$, $5.35$, and $3.50$~dB, respectively. Increasing $K$ beyond $400$ produces relatively small changes in the received SNR for \textsc{P-PGNBB}, which remains between $53.50$ and $54.09$~dB.

	To compare measurement requirements at a common performance level, we consider a received SNR of $53$~dB. As can be seen, \textsc{P-PGNBB}, \textsc{CSM}, \textsc{PGNN}, and \textsc{BORN} first reach this level at $K=400$, $1600$, $3200$, and $12800$, respectively. Thus, \textsc{P-PGNBB} uses $25\%$, $12.5\%$, and approximately $3.1\%$ of the number of measurements required by \textsc{CSM}, \textsc{PGNN}, and \textsc{BORN}, respectively, to reach this threshold in the experiment. The advantage over \textsc{PGNN} is attributable to the compact channel representation, i.e., for $N=256$ and $L=3$, \textsc{P-PGNBB} learns only $14$ real parameters, whereas \textsc{PGNN} learns up to $514$.

	Fig.~\ref{fig:hardware_alg_benchmark}(b) investigates the impact of the assumed model order on received SNR. At $K=400$, the models with $L=1$ and $L=2$ achieve a received SNR of only $35.80$ and $38.05$~dB, respectively, while using $L=3$ results in $53.50$~dB. With $K=1600$, all three models achieve similar received SNRs of $53.35$, $53.37$, and $53.70$~dB for $L=1$, $2$, and $3$, respectively. These results indicate that the lower-order models can also produce effective RIS configurations when more measurements are available, although setting $L=3$ reaches this performance level with a smaller number of measurements.
	%-------------------------------------------------------------------------------------
	\subsubsection{Learning behavior analysis}
	\label{subse:results_learning_behavior}

	\begin{figure}[t]
		\centering
		\includegraphics[width=\linewidth]{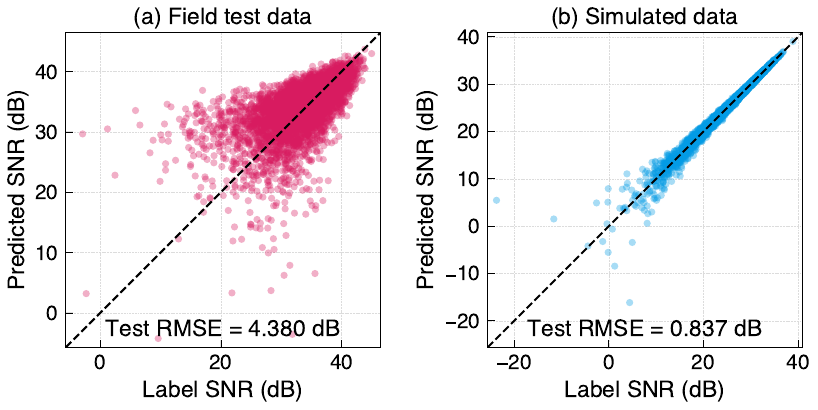}
		\caption{Predicted SNR versus reference SNR on test data after training \textsc{P-PGNBB}: (a) field-test data with $K=400$ and $L=3$ and (b) simulated data with $K=400$ and $L_{\mathrm{true}}=L=2$. The dashed line denotes perfect prediction.}
		\label{fig:predict_vs_label}
	\end{figure}

	Fig.~\ref{fig:predict_vs_label} examines how well the trained \textsc{P-PGNBB} model predicts the SNR of unseen RIS configurations using (a) field-test data with $K=400$ training measurements and $L=3$ and (b) simulated data with $K=400$ and $L_{\mathrm{true}}=L=2$. Each simulation uses $5000$ test samples. The simulated predictions are concentrated closely around the perfect-prediction line, resulting in a root mean square error (RMSE) of $0.837$~dB. The field-test predictions exhibit a wider spread but retain a positive association with the measured labels, with a test RMSE of $4.380$~dB. Receiver noise, modeling mismatch, and temporal variations may contribute to the larger field-test error, although their individual contributions are not isolated in this experiment.

	\begin{figure}[t]
		\centering
		\includegraphics[width=0.9\linewidth]{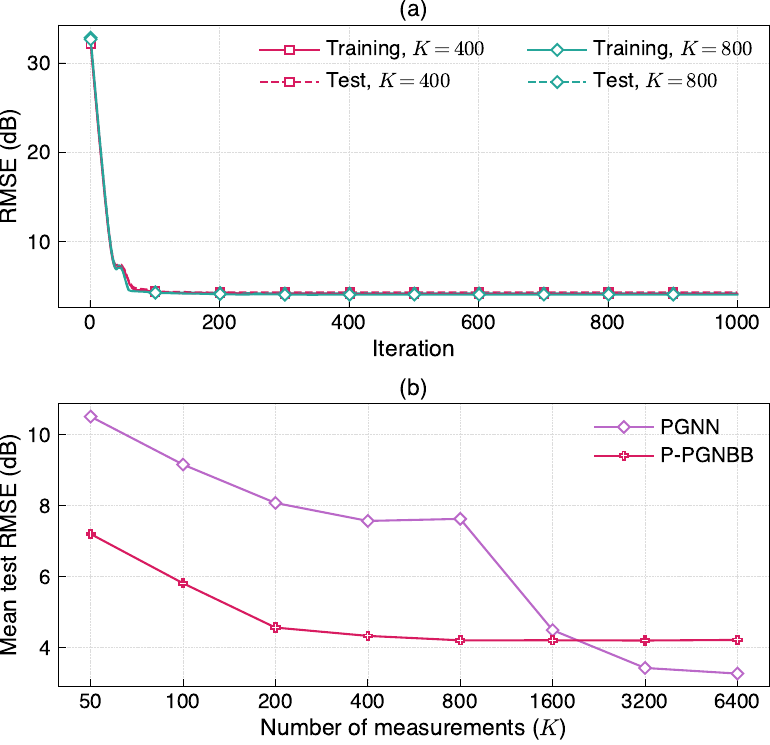}
		\vspace{8pt}
		\caption{Learning efficiency on the field-test data: (a) training and test RMSE of \textsc{P-PGNBB} versus iteration number for $K = 400, 800$; and (b) mean test RMSE of \textsc{PGNN} and \textsc{P-PGNBB} versus $K$. Herein, $L = 3$.}
		\label{fig:learning_efficiency}
	\end{figure}
	
	\begin{figure}[t]
		\centering
		\includegraphics[width=\linewidth]{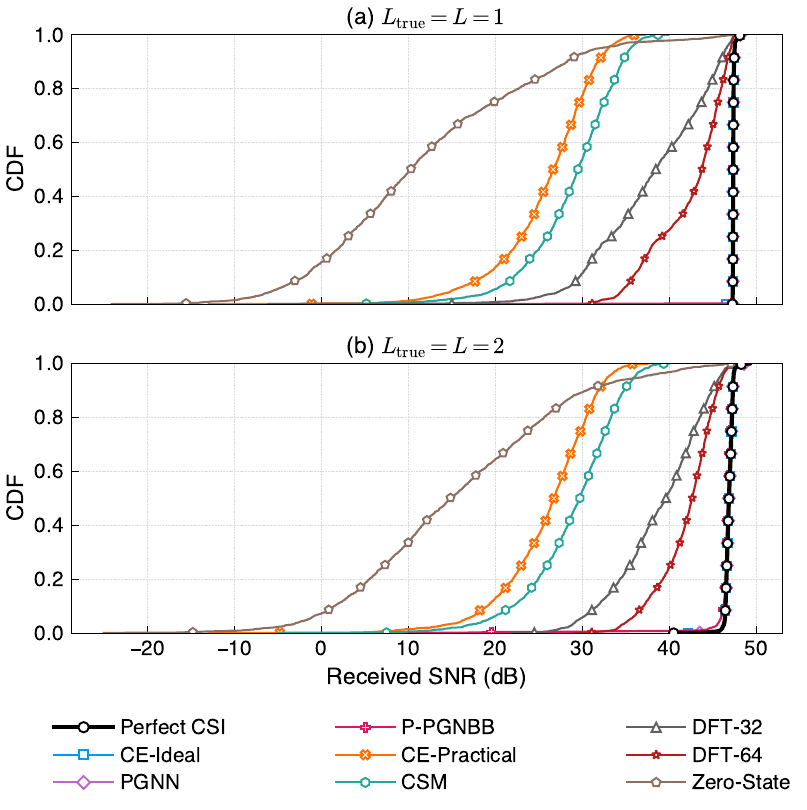}
		\caption{The cumulative distribution function (CDF) of the received SNR under the considered algorithms. In this simulation, we use a $10\times10$ RIS with $Q=4$ in two cases: (a) $L_{\mathrm{true}}=L=1$ and (b) $L_{\mathrm{true}}=L=2$.}
		\label{fig:algorithm_benchmark_q4_cdf}
	\end{figure}
	Next, we investigate the learning efficiency of \textsc{P-PGNBB} in terms of optimizer convergence and the number of training measurements. Fig.~\ref{fig:learning_efficiency}(a) shows the achieved RMSE over the training iterations for two cases: $K = 400, 800$. As can be seen, most of the reduction in training and test RMSE occurs within approximately the first $100$ $\mathtt{L\mbox{-}BFGS}$ iterations, followed by smaller changes as the curves approach stable values. For both values of $K$, the training and test curves remain close, indicating a small generalization gap for these subsets of the recorded acquisition. At the final iteration, the mean test RMSE is $4.333$~dB for $K=400$ and $4.209$~dB for $K=800$. Thus, doubling the number of training measurements reduces the mean test RMSE by $0.124$~dB. These results indicate that doubling the number of online measurements in this case does not improve the training significantly. This result is aligned with the insights studied in Fig. \ref{fig:hardware_alg_benchmark}(a).

	Fig.~\ref{fig:learning_efficiency}(b) compares the mean test RMSE of \textsc{PGNN} and \textsc{P-PGNBB} for different numbers of training measurements.\footnote{Note that only \textsc{PGNN} and \textsc{P-PGNBB} are included because both methods are based on neural-network training, whereas the other benchmark algorithms do not produce a model whose prediction error can be evaluated on a test set.} The results are averaged over ten independent random draws of the training data. When $K$ is low, \textsc{P-PGNBB} reduces the mean test RMSE more rapidly than \textsc{PGNN}. At $K=50$, its mean test RMSE is $7.204$~dB, compared with $10.512$~dB for \textsc{PGNN}. At $K=400$, the corresponding values are $4.333$ and $7.572$~dB. This difference is consistent with the compact parametric representation of \textsc{P-PGNBB}, which contains $4L+2=14$ trainable real parameters for $L=3$, compared with $2N+2=514$ parameters for \textsc{PGNN}. However, the two methods achieve similar errors at $K=1600$, and \textsc{PGNN} attains lower mean test RMSEs at $K=3200$ and $6400$\basicfix{.} These results indicate that the physics-guided parameterization improves data efficiency at smaller measurement budgets, whereas the more general model can achieve lower prediction error when substantially more measurements are available.
	%-------------------------------------------------------------------------------------

	\subsection{Simulation Setup}
	\label{subse:simulation_setup}
	
	We use simulations to benchmark \textsc{P-PGNBB} and the considered baseline methods against the upper bound. In the simulation setup, the TX--RIS, RIS--RX, and direct TX--RX link distances are all set to $15$ m. We set $P=30$ dBm. Unless stated otherwise, the results are analyzed over $2000$ independent channel realizations.
	
	\begin{figure}[t]
		\centering
		\includegraphics[width=\linewidth]{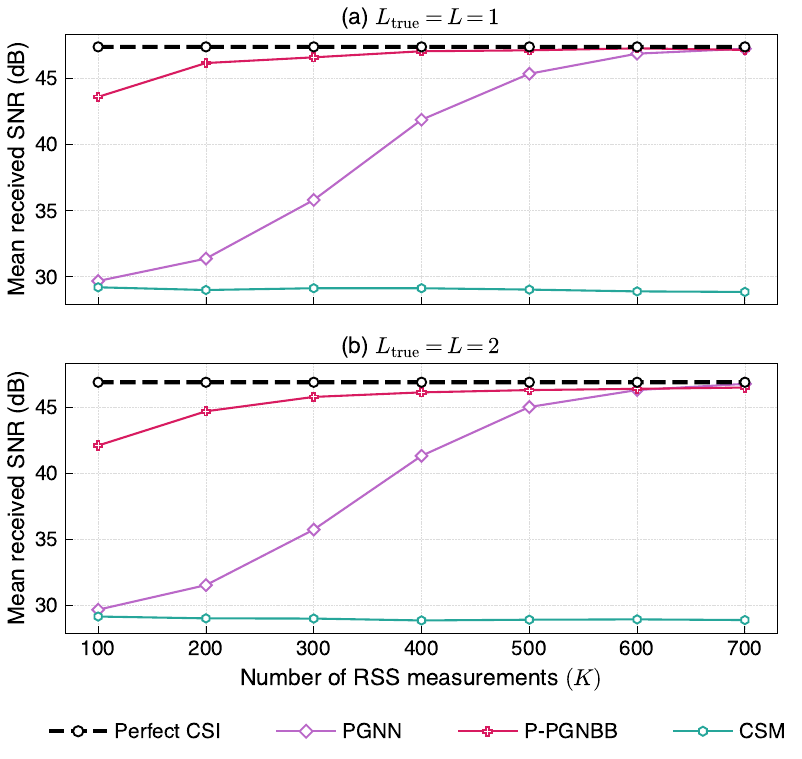}
		\caption{Average received SNR versus the number of RSS measurements for a $10\times10$ RIS with $Q=4$: (a) $L_{\mathrm{true}}=L=1$ and (b) $L_{\mathrm{true}}=L=2$.}
		\label{fig:rss_measurement_efficiency}
	\end{figure}
	%-------------------------------------------------------------------------------------
	\subsubsection{Channel models}
	
	Each channel is generated from the far-field parametric model shown in Section~\ref{subse:parametric_channel_model}. The TX-to-RIS and RIS-to-RX azimuth and elevation angles are drawn independently and uniformly from $[-60^\circ,60^\circ]$. We generate channels containing $L_{\mathrm{true}}\in\{1,2,3, 4\}$ components. For each realization, the relative component powers and phases are generated as
	\begin{equation}
		\begin{aligned}
			u_\ell&\sim\operatorname{Exp}(1),
			&
			w_\ell&=\frac{u_\ell}{\sum_{j=1}^{L_{\mathrm{true}}}u_j},
			\\
			\varphi_\ell&\sim\mathcal U[0,2\pi),
			&
			\bar{\gamma}_\ell&=\sqrt{w_\ell}e^{\mathrm{j}\varphi_\ell}.
		\end{aligned}
		\label{eq:simulation_relative_path_gains}
	\end{equation}
	
	The large-scale path-loss model follows~\cite{dinhvanTWC2026BORN}. Specifically, for each RIS-side link $s\in\{\mathrm{TR},\mathrm{RR}\}$, the path loss is modeled as $\mathrm{PL}_{s}(d_s)\,[\mathrm{dB}] = 30 + 22\log_{10}(d_s)$, where $d_s$ is the distance in meters. The noise power is set as $-90$ dBm. We consider the most challenging case where the background channel is absent, i.e., $h_0 = 0$.

	We set $\omega_k=0$ for CE-Ideal, corresponding to phase-coherent measurements. For CE-Practical, the channel follows the CFO model in~\eqref{eq:cfo_phase_model}, with $\delta_f=100$ Hz, $t_k=(k-1)$ ms, and an initial phase $\theta_0\sim\mathcal{U}[0,2\pi)$ independently generated for each channel realization. Both benchmarks employ a ZC pilot sequence of length $L_p=1023$. As described in Section~\ref{subse:ideal_channel_estimation}, $N+1$ RIS training configurations are applied sequentially to estimate the background and cascaded channels.
	%-------------------------------------------------------------------------------------
	\subsection{Simulation Results}
	\subsubsection{Algorithm benchmark under Q-Ary RIS}

	\begin{figure}[!t]
		\centering
		\includegraphics[width=0.9 \linewidth]{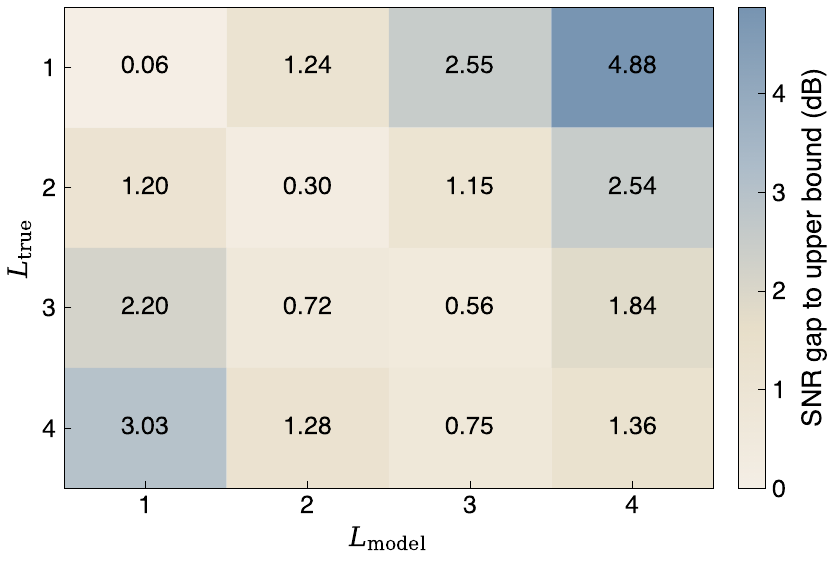}
		\vspace{5pt}
		\caption{Mean received-SNR gap between \textsc{P-PGNBB} and the upper bound achieved under different $L_\mathrm{true}$ and $L_\mathrm{model}$. Here, we use \basicfix{a} $10\times10$ RIS with $Q=4$ and $K=500$. Smaller values indicate better performance.}
		\label{fig:model_order_mismatch}
	\end{figure}

	We first evaluate the received-SNR performance of the considered algorithms. Fig.~\ref{fig:algorithm_benchmark_q4_cdf} shows the CDF of the received SNR for a $10\times10$ RIS with $Q=4$ and $L=L_{\mathrm{true}}\in\{1,2\}$. For a fair comparison among the RSS-based methods, \textsc{PGNN}, \textsc{P-PGNBB}, and \textsc{CSM} use the same $K=2000$ random RIS configurations. As can be seen, \textsc{P-PGNBB}, \textsc{PGNN}, and CE-Ideal closely approach the upper bound, i.e., Perfect CSI, for both channel orders. This shows that the proposed low-dimensional physical representation can recover RIS configurations whose performance is close to that obtained with perfect channel knowledge. In this simulation, CE-Ideal performs well because the complex pilot observations remain phase coherent across RIS configurations. In contrast, CE-Practical suffers a substantial performance loss due to the effects of CFO. Specifically, the pilot-based channel estimator combines sequential observations as though they shared a common phase reference, resulting in inconsistent estimates of the element-wise channel phases and, consequently, a poor RIS configuration. Its median received SNR decreases to $26.459$ dB for $L_{\mathrm{true}}=1$ and $26.686$ dB for $L_{\mathrm{true}}=2$, which are approximately $20.9$ and $20.2$ dB below that of Perfect CSI. By contrast, \textsc{PGNN}, \textsc{P-PGNBB}, and \textsc{CSM} rely on RSS measurements and are therefore less influenced by the CFO effect.

	The DFT codebooks provide reasonable performance without explicitly fitting a channel model. However, their finite angular resolution limits the likelihood of selecting the best RIS configuration. Increasing the codebook size from $32$ to $64$ improves the median received SNR from $37.767$ to $43.939$ dB for $L_{\mathrm{true}}=1$, and from $39.445$ to $42.620$ dB for $L_{\mathrm{true}}=2$. Nevertheless, DFT-64 remains $3.431$ and $4.259$ dB below the corresponding Perfect CSI medians. Among the RIS configuration algorithms, \textsc{CSM} yields the weakest performance, with median received SNRs of $29.529$ and $29.579$ dB for $L_{\mathrm{true}}=1$ and $L_{\mathrm{true}}=2$, respectively. As expected, the zero-state approach obtains the lowest overall received SNR.
	
	%-------------------------------------------------------------------------------------
	
	\subsubsection{Measurement efficiency benchmark}
	We next investigate how the number of online RSS measurements affects the received SNR performance. Fig.~\ref{fig:rss_measurement_efficiency} compares the average received SNR under \textsc{P-PGNBB}, \textsc{PGNN}, and \textsc{CSM} for different values of $K$. It can be clearly seen that \textsc{P-PGNBB} is substantially more measurement-efficient than \textsc{PGNN}. At $K=100$, \textsc{P-PGNBB} achieves average received SNRs of $43.60$ and $42.43$ dB for $L_{\mathrm{true}}=1$ and $L_{\mathrm{true}}=2$, respectively, compared with $29.15$ and $29.75$ dB for \textsc{PGNN}. At $K=200$, the gaps between \textsc{P-PGNBB} and perfect CSI are only $1.74$ and $1.10$ dB for $L_{\mathrm{true}}=1$ and $L_{\mathrm{true}}=2$, respectively. In comparison, the corresponding gaps for \textsc{PGNN} remain $15.20$ and $16.58$ dB. This is because \textsc{PGNN} must learn $2N+2$ real parameters, whereas \textsc{P-PGNBB} learns only $2+4L$ real parameters. Both methods approach the perfect-CSI performance at approximately $K=600$. However, \textsc{P-PGNBB} already achieves a small performance gap at $K=200$, requiring only one-third of the measurements needed by \textsc{PGNN} to reach a comparable SNR. In contrast, the received SNR achieved by \textsc{CSM} remains at approximately $29$~dB across the considered values of $K$, indicating that increasing the measurement budget alone does not improve its performance in this setting. This behavior is consistent with the limitations of \textsc{CSM} when the background channel is weak, as shown in \cite{dinhvanTWC2026BORN}.
	\begin{table}[!t]
		\caption{Average total processing time of \textsc{P-PGNBB} in ms}
		\label{tab:processing_time_total}
		\centering
		\renewcommand{\arraystretch}{1.10}
		\setlength{\tabcolsep}{5pt}
		\begin{tabular*}{\columnwidth}{@{\extracolsep{\fill}}lrrrrr@{}}
			\toprule
			Method & $L=1$ & $L=2$ & $L=3$ & $L=4$ & $L=5$ \\
			\midrule
			%\rowcolor{gray!12}
			\textbf{\textsc{P-PGNBB}} & 47.5 & 52.7 & 53.2 & 52.8 & 54.3 \\
			\bottomrule
		\end{tabular*}
	\end{table}
	\begin{table}[!t]
		\caption{Average processing-time breakdown over in ms}
		\label{tab:processing_time_breakdown}
		\centering
		\renewcommand{\arraystretch}{1.12}
		\setlength{\tabcolsep}{3.5pt}
		\begin{tabular*}{\columnwidth}{@{\extracolsep{\fill}}lrrr@{}}
			\toprule
			Method & Estimation/training & RIS optimization & Total \\
			\midrule
			CE-Ideal             & 2.26 & 1.33 & 3.59 \\
			\textsc{PGNN}        & 35.1 & 1.39 & 36.5 \\
			\rowcolor{gray!12}
			\textbf{\textsc{P-PGNBB}} & 50.7 & 1.40 & 52.1 \\
			\textsc{BORN}        & 375  & 2.96 & 378  \\
			\textsc{CSM}         & N/A  & N/A  & 0.143 \\
			DFT-32               & N/A  & N/A  & 0.00241 \\
			DFT-64               & N/A  & N/A  & 0.00157 \\
			\bottomrule
		\end{tabular*}
	\end{table}
	%-------------------------------------------------------------------------------------
	\subsubsection{Robustness to model-order mismatch}

	We next examine the sensitivity of \textsc{P-PGNBB} to the selected model order $L_{\mathrm{model}}$. In practice, the true channel order $L_{\mathrm{true}}$ is unknown, so $L_{\mathrm{model}}$ must be selected without this information. Fig.~\ref{fig:model_order_mismatch} shows the average received-SNR gap between \textsc{P-PGNBB} and the upper bound achieved via Perfect CSI for $L_{\mathrm{true}},L_{\mathrm{model}}\in\{1,2,3,4\}$. When the model order is matched, all gaps remain below $1.5$~dB, demonstrating the effectiveness of the proposed method across the considered channel orders. The largest losses occur under severe mismatch, e.g., the gaps reach $4.88$~dB for $(L_{\mathrm{true}},L_{\mathrm{model}})=(1,4)$, and $3.03$~dB for $(4,1)$. This is because the true channel model becomes substantially different from the assumed model as the model-order mismatch increases. In contrast, when $|L_{\mathrm{true}}-L_{\mathrm{model}}|<2$, the gap remains below $2$~dB, indicating an acceptable performance loss under small model-order mismatch.
%-------------------------------------------------------------------------------------
	\subsubsection{Computational latency benchmark}
	
	We next examine the computational time for a $10\times10$ RIS with $Q=2$. For \textsc{P-PGNBB}, we use $K=200$ and $L\in\{1,\ldots,5\}$.\footnote{The implementation was benchmarked on a commodity laptop. Therefore, these results are intended primarily to provide relative comparisons. With dedicated hardware and an optimized implementation, the processing time could be further reduced.} Since the other algorithms do not use $L$ as a model parameter, Table~\ref{tab:processing_time_total} reports only the total processing time of \textsc{P-PGNBB} for each $L$. In comparison, Table~\ref{tab:processing_time_breakdown} reports the processing times of all considered algorithms, averaged over the tested values of $L$, with a breakdown into estimation/training and RIS optimization where applicable. For \textsc{CSM} and the DFT codebook methods, these stages are marked as N/A and only the total configuration-processing time is reported.
	
	As shown in Table~\ref{tab:processing_time_total}, the processing time of \textsc{P-PGNBB} increases moderately from $47.5$ to $54.3$~ms as $L$ increases from $1$ to $5$. Its average total processing time is $52.1$~ms, compared with $36.5$~ms for \textsc{PGNN}, $378$~ms for \textsc{BORN}, and $3.59$~ms for CE-Ideal. By contrast, \textsc{CSM} and the DFT codebook methods require only $0.143$~ms and approximately $0.002$~ms, respectively. Furthermore, Table~\ref{tab:processing_time_breakdown} shows that estimation or training dominates the processing time of the model-based methods, whereas the subsequent binary RIS optimization requires only $1.33$--$2.96$~ms. Thus, the main computational burden of \textsc{P-PGNBB} and \textsc{PGNN} comes from neural-network training. This processing cost is the trade-off for the reduced number of online measurements demonstrated in Fig.~\ref{fig:rss_measurement_efficiency}.
%-------------------------------------------------------------------------------------
	\section{Conclusion}
	\label{se:conclusion}
	This paper studied RIS configuration using RSS-based channel estimation. In particular, hardware measurements revealed substantial phase drift across successive pilot blocks, which can degrade pilot-based channel estimation when the observations are combined without compensation. By contrast, RSS measurements are significantly less influenced by these phase rotations and therefore avoid the need for explicit CFO estimation and compensation. To improve measurement efficiency within this RSS-based framework, we developed \textsc{P-PGNBB}, which learns a compact effective cascaded channel using an $L$-component far-field representation. Specifically, it is a shallow physics-guided neural network with a small number of trainable parameters. Simulations show that the proposed method approached the received SNR obtained with perfect CSI and outperformed the considered model-free baseline. Furthermore, it required only one-third of the measurements needed by \textsc{PGNN} to achieve comparable received SNR in the evaluated scenarios. Hardware experiments also demonstrated comparable received SNR to existing methods with substantially fewer RSS measurements.

	\appendices
	\section{P-PGNBB Initialization Approach}
	\label{app:ppgnbb_initialization}
	This appendix details the initialization procedure summarized in Algorithm~\ref{alg:ppgnbb_initialization}.

	\subsection{\texorpdfstring{Single-Component Grid Search for \(L=1\)}{Single-Component Grid Search for L = 1}}

	For $L=1$, we set $h_0^{(0)}=0$ and form a uniform grid $\mathcal G=\mathcal G_y\times\mathcal G_z$. For each candidate $\boldsymbol{\kappa}\in\mathcal G$, we define
	\begin{equation}
		a_k(\boldsymbol{\kappa})
		\triangleq
		\rho
		\abs{\bm{x}_k^\top\bm{b}(\boldsymbol{\kappa})}^{2}.
		\label{eq:grid_candidate_snr_response}
	\end{equation}
	The gain amplitude for each candidate is obtained by least squares. In particular, scaling $a_k(\boldsymbol{\kappa})$ by $\abs{\gamma^{(0)}(\boldsymbol{\kappa})}^2$ and fitting the SNR labels gives
	\begin{equation}
		\abs{\gamma^{(0)}(\boldsymbol{\kappa})}
		=
		\left(
		\frac{
			\sum_{k=1}^{K}
			r_k
			a_k(\boldsymbol{\kappa})
		}{
			\sum_{k=1}^{K}
			a_k^2(\boldsymbol{\kappa})
		}
		\right)^{1/2}.
		\label{eq:grid_gain_amplitude}
	\end{equation}
	The fitting error associated with candidate $\boldsymbol{\kappa}$ is expressed as
	\begin{equation}
		\mathcal J_1(\boldsymbol{\kappa})
		\triangleq
		\frac{1}{K}
		\sum_{k=1}^{K}
		\left[
		r_k
		-
		\abs{\gamma^{(0)}(\boldsymbol{\kappa})}^{2}
		a_k(\boldsymbol{\kappa})
		\right]^2.
		\label{eq:grid_candidate_snr_error}
	\end{equation}
	The spatial-frequency initialization is therefore selected as
	\begin{equation}
		\boldsymbol{\kappa}^{(0)}
		\in
		\underset{\boldsymbol{\kappa}\in\mathcal G}{\argmin}\;
		\mathcal J_1(\boldsymbol{\kappa}).
		\label{eq:grid_frequency_initialization}
	\end{equation}
	Note that only the gain magnitude is identifiable when $L=1$ and $h_0^{(0)}=0$. Hence, after selecting $\boldsymbol{\kappa}^{(0)}$, we choose the gain initialization as $\gamma^{(0)}\triangleq\abs{\gamma^{(0)}(\boldsymbol{\kappa}^{(0)})}$, which is real and nonnegative.

	\subsection{\texorpdfstring{Multi-Component Grid Search for \(L>1\)}{Multi-Component Grid Search for L > 1}}

	For $L>1$, we first apply the procedure in~\eqref{eq:grid_candidate_snr_response}--\eqref{eq:grid_frequency_initialization} to the original labels $\{r_k\}_{k=1}^{K}$ to obtain $\boldsymbol{\kappa}_1^{(0)}$ and its provisional gain $\widetilde\gamma_1^{(0)}$. Suppose that $\ell-1$ components have been selected for some $\ell\in\{2,\ldots,L\}$. For the $k$-th probing configuration, define $z_{kj}^{(0)}\triangleq\bm{x}_k^\top\bm{b}(\boldsymbol{\kappa}_j^{(0)})$. Using the provisional gains from the previous scans, the predicted SNR is
	\begin{equation}
		\widehat r_k^{(\ell-1)}
		=
		\rho
		\abs{
		\sum_{j=1}^{\ell-1}
		\widetilde\gamma_j^{(0)}
		z_{kj}^{(0)}
		}^{2}.
		\label{eq:multicomponent_initial_prediction}
	\end{equation}
	The corresponding nonnegative residual label is
	\begin{equation}
		r_{k,\mathrm{res}}^{(\ell)}
		\triangleq
		\max
		\left\{
		r_k-\widehat r_k^{(\ell-1)},0
		\right\}.
		\label{eq:multicomponent_initial_residual}
	\end{equation}
	We then reapply~\eqref{eq:grid_candidate_snr_response}--\eqref{eq:grid_frequency_initialization} to the residual labels $\{r_{k,\mathrm{res}}^{(\ell)}\}_{k=1}^{K}$ to obtain $\boldsymbol{\kappa}_\ell^{(0)}$ and its provisional gain $\widetilde\gamma_\ell^{(0)}$. We update~\eqref{eq:multicomponent_initial_prediction} and~\eqref{eq:multicomponent_initial_residual} and repeat this procedure until all $L$ spatial frequencies have been selected.

	Because the provisional gains are obtained from separate residual scans, they may not accurately capture the coherent interactions among components. Once all $L$ spatial frequencies have been selected, we therefore refit the complex gains jointly using the original SNR labels. Define
	\begin{equation}
		\bm\gamma
		\triangleq
		[\gamma_1,\ldots,\gamma_L]^\top,
		\qquad
		\bm G
		\triangleq
		\bm\gamma\bm\gamma^{\mathrm{H}}.
		\label{eq:multicomponent_lifted_gain_matrix}
	\end{equation}
	For the $k$-th probing configuration, using $G_{jm}=\gamma_j\gamma_m^*$, the coherent SNR prediction becomes
	\begin{equation}
	\begin{aligned}
		\widehat r_k(\bm\gamma)
		&\triangleq
		\rho
		\abs{
		\sum_{j=1}^{L}
		\gamma_j z_{kj}^{(0)}
		}^{2} =
		\rho
		\Biggl[
		\sum_{j=1}^{L}
		G_{jj}\abs{z_{kj}^{(0)}}^2\\
		&+
		2
		\sum_{j=1}^{L-1}
		\sum_{m=j+1}^{L}
		\Real{
		G_{jm}
		z_{kj}^{(0)}
		\bigl(z_{km}^{(0)}\bigr)^*
		}
		\Biggr].
	\end{aligned}
	\label{eq:multicomponent_lifted_prediction}
	\end{equation}
	Denote the final expression by $\widehat r_k(\bm G)$. Because this expression is linear in the entries of the Hermitian matrix $\bm G\in\C^{L\times L}$, we fit $\bm G$ to the original SNR labels by solving
	\begin{equation}
		\widetilde{\bm G}
		\in
		\underset{\bm G=\bm G^{\mathrm{H}}}{\argmin}
		\frac{1}{K}
		\sum_{k=1}^{K}
		\left[
		r_k-\widehat r_k(\bm G)
		\right]^2.
		\label{eq:multicomponent_hermitian_fit}
	\end{equation}
	Ideally, $\bm G=\bm\gamma\bm\gamma^{\mathrm{H}}$ is positive semidefinite and rank one. However, noise and the unconstrained fit in~\eqref{eq:multicomponent_hermitian_fit} may prevent $\widetilde{\bm G}$ from having this structure. Let $(\lambda_{\max},\bm v_{\max})$ denote the dominant eigenpair of $\widetilde{\bm G}$. Its positive rank-one approximation yields
	\begin{equation}
		\bm\gamma^{(0)}
		=
		\sqrt{\max\{\lambda_{\max},0\}}\,
		\bm v_{\max}.
		\label{eq:multicomponent_gain_initialization}
	\end{equation}

	\section{Derivation of the Q-Ary RIS Optimization}
	\label{app:qary_optimization_derivation}
	This appendix derives the directional event sweep used in Algorithm~\ref{alg:qary_event_sweep}. Define
	$\widehat g(\bm{x})\triangleq\widehat h_0+\bm{x}^{\top}\widehat{\bm h}$. Using
	$\abs{z}=\max_{\phi\in[0,2\pi)}\Real{e^{-\mathrm{j}\phi}z}$ and interchanging the two maximizations, maximizing $\abs{\widehat g(\bm{x})}$ over $\bm{x}\in\mathcal X_Q^N$ is equivalent to
	\begin{equation}
		\underset{\mathclap{\phi\in[0,2\pi)}}{\max} \ 
		\left[
		\Real{e^{-\mathrm{j}\phi}\widehat h_0}
		+
		\sum_{n=1}^{N} \ \
		\underset{\mathclap{x_n\in\mathcal X_Q}}{\max} \
		\Real{e^{-\mathrm{j}\phi}\widehat h_n x_n}
		\right].
		\label{eq:qary_directional_reformulation}
	\end{equation}
	For fixed $\phi$, the first term is independent of $\bm x$, while the $n$-th inner maximization depends only on $x_n$. Hence, the maximizing RIS configuration is obtained elementwise as
	\begin{equation}
		x_n(\phi)
		\in
		\underset{x\in\mathcal X_Q}{\argmax} \ 
		\Real{e^{-\mathrm{j}\phi}\widehat h_nx}.
		\label{eq:qary_projection_selection}
	\end{equation}
	Eq.~\eqref{eq:qary_projection_selection} selects the available phase state that places $\widehat h_nx_n$ closest to the direction $\phi$. As $\phi$ varies, this state remains unchanged between the midpoints of adjacent contribution phases and can switch only at
	\begin{equation}
		\begin{gathered}
			\vartheta_{n,q}
			=
			\left(
			\arg(\widehat h_n)
			+
			\frac{(2q+1)\pi}{Q}
			\right)
			\bmod 2\pi,\\
			q=0,\ldots,Q-1.
		\end{gathered}
		\label{eq:qary_phase_boundaries}
	\end{equation}
	Between consecutive distinct boundaries, every $x_n(\phi)$ and hence $\bm{x}(\phi)$ remain unchanged. Therefore, evaluating one configuration per interval exhausts all candidates needed to maximize~\eqref{eq:qary_directional_reformulation}. Algorithm~\ref{alg:qary_event_sweep} performs precisely this sweep and thus returns a globally optimal solution of~$\mathrm{(P2)}$.

	\providecommand{\BIBdecl}{\interlinepenalty=10000}
	\bibliographystyle{IEEEtran}
	\bibliography{bibfile_gen}

@article{laiTMC2025adaptiveBlind,
	author  = {Wenhai Lai and Wenyu Wang and Fan Xu and Xin Li and Shaobo Niu and Kaiming Shen},
	title   = {Adaptive Blind Beamforming for Intelligent Surface},
	journal = {IEEE Trans. Mobile Comput.},
	year    = {2025},
	volume  = {24},
	number  = {2},
	pages   = {907--923},
	doi     = {10.1109/TMC.2024.3468618},
	month = {Feb.}
}

@misc{tmytekWeb2025xrifle,
	author       = {{TMYTEK}},
	title        = {{XRifle} Dynamic {RIS}},
	note         = {{Accessed: Oct. 16, 2026}},
	url = {https://tmytek.com/products/components/xrifle-dynamic-ris}
}

@article{peiTCOM2021RISPrototype,
	author  = {X. Pei and H. Yin and L. Tan and L. Cao and Z. Li and K. Wang and K. Zhang and E. Bj{\"o}rnson},
	title   = {{RIS}-Aided Wireless Communications: Prototyping, Adaptive Beamforming, and Indoor/Outdoor Field Trials},
	journal = {IEEE Trans. Commun.},
	year    = {2021},
	volume  = {69},
	number  = {12},
	pages   = {8627--8640},
	doi     = {10.1109/TCOMM.2021.3116151},
	month = {Dec.}
}

@inproceedings{arunNSDI2020RFocus,
	author    = {V. Arun and H. Balakrishnan},
	title     = {{RFocus}: Beamforming Using Thousands of Passive Antennas},
	booktitle = {Proc. 17th USENIX Symp. Netw. Syst. Design Implement. (NSDI)},
	pages     = {1047--1061},
	year      = {2020},
	month = {Feb.}
}

@article{wuTWC2019jointBeamforming,
	author  = {Q. Wu and R. Zhang},
	journal = {IEEE Trans. Wireless Commun.},
	title   = {Intelligent Reflecting Surface Enhanced Wireless Network via Joint Active and Passive Beamforming},
	year    = {2019},
	volume  = {18},
	number  = {11},
	pages   = {5394--5409},
	doi     = {10.1109/TWC.2019.2936025},
	month = {Nov.}
}

@article{diRenzoJSAC2020smartRadio,
	author  = {Di Renzo, M. and A. Zappone and M. Debbah and M.-S. Alouini and C. Yuen and J. de Rosny and S. Tretyakov},
	journal = {IEEE J. Sel. Areas Commun.},
	title   = {Smart Radio Environments Empowered by Reconfigurable Intelligent Surfaces: How It Works, State of Research, and The Road Ahead},
	year    = {2020},
	volume  = {38},
	number  = {11},
	pages   = {2450--2525},
	doi     = {10.1109/JSAC.2020.3007211},
	month = {Nov.}
}

@article{liuCOMST2021RISSurvey,
	author  = {Y. Liu and X. Liu and X. Mu and T. Hou and J. Xu and Di Renzo, M. and N. Al-Dhahir},
	journal = {IEEE Commun. Surveys Tuts.},
	title   = {Reconfigurable Intelligent Surfaces: Principles and Opportunities},
	year    = {2021},
	volume  = {23},
	number  = {3},
	pages   = {1546--1577},
	doi     = {10.1109/COMST.2021.3077737},
	month = {3rd Quart.}
}

@article{basarAccess2019RIS,
	author  = {E. Basar and Di Renzo, M. and J. de Rosny and M. Debbah and M.-S. Alouini and R. Zhang},
	journal = {IEEE Access},
	title   = {Wireless Communications Through Reconfigurable Intelligent Surfaces},
	year    = {2019},
	volume  = {7},
	pages   = {116753--116773},
	doi     = {10.1109/ACCESS.2019.2935192},
	month = {Aug.}
}

@ARTICLE{dinhvanJIOT2022,
	author  = {Dinh-Van, Son and Hoang, Tiep M. and Trestian, Ramona and Nguyen, Huan X.},
	title   = {Unsupervised Deep-Learning-Based Reconfigurable Intelligent Surface-Aided Broadcasting Communications in Industrial {IoTs}},
	journal = {IEEE Internet Things J.},
	year    = {2022},
	volume  = {9},
	number  = {19},
	pages   = {19515--19528},
	doi     = {10.1109/JIOT.2022.3169276},
	month = {Oct.}
}

@article{wuTCOM2021RISTutorial,
	author  = {Qingqing Wu and Shuowen Zhang and Beixiong Zheng and Changsheng You and Rui Zhang},
	title   = {Intelligent Reflecting Surface-Aided Wireless Communications: A Tutorial},
	journal = {IEEE Trans. Commun.},
	year    = {2021},
	volume  = {69},
	number  = {5},
	pages   = {3313--3351},
	doi     = {10.1109/TCOMM.2021.3051897},
	month = {May}
}

@article{youJSAC2020channelEstimation,
	author  = {Changsheng You and Beixiong Zheng and Rui Zhang},
	title   = {Channel Estimation and Passive Beamforming for Intelligent Reflecting Surface: Discrete Phase Shift and Progressive Refinement},
	journal = {IEEE J. Sel. Areas Commun.},
	year    = {2020},
	volume  = {38},
	number  = {11},
	pages   = {2604--2620},
	doi     = {10.1109/JSAC.2020.3007056},
	month = {Nov.}
}

@article{zhengCOMST2022channelEstimationSurvey,
	author  = {Beixiong Zheng and Changsheng You and Weidong Mei and Rui Zhang},
	title   = {A Survey on Channel Estimation and Practical Passive Beamforming Design for Intelligent Reflecting Surface Aided Wireless Communications},
	journal = {IEEE Commun. Surveys Tuts.},
	year    = {2022},
	volume  = {24},
	number  = {2},
	pages   = {1035--1071},
	doi     = {10.1109/COMST.2022.3155305},
	month = {2nd Quart.}
}

@article{renTWC2023blindBeamforming,
	author  = {Shuyi Ren and Kaiming Shen and Yaowen Zhang and Xin Li and Xin Chen and Zhi-Quan Luo},
	title   = {Configuring Intelligent Reflecting Surface With Performance Guarantees: Blind Beamforming},
	journal = {IEEE Trans. Wireless Commun.},
	year    = {2023},
	volume  = {22},
	number  = {5},
	pages   = {3355--3370},
	doi     = {10.1109/TWC.2022.3217679},
	month = {May}
}

@inproceedings{sunGLOBECOM2023PGNN,
	author    = {He Sun and Weidong Mei and Lipeng Zhu and Rui Zhang},
	title     = {User Power Measurement Based {IRS} Channel Estimation via Single-Layer Neural Network},
	booktitle = {Proc. IEEE Global Commun. Conf. (GLOBECOM)},
	year      = {2023},
	pages     = {6579--6584},
	doi       = {10.1109/GLOBECOM54140.2023.10437096},
	month = {Dec.}
}

@article{dinhvanTWC2026BORN,
	author  = {Son Dinh-Van and Nam Phuong Tran and Matthew D. Higgins},
	title   = {Near-Optimal Reconfigurable Intelligent Surface Configuration: Blind Beamforming With Sensing},
	journal = {IEEE Trans. Wireless Commun.},
	year    = {2026},
	volume  = {25},
	pages   = {22572--22587},
	doi     = {10.1109/TWC.2026.3726665},
	month = {Aug.}
}

@article{haghshenasTWC2024parametricCE,
	author  = {Mehdi Haghshenas and Parisa Ramezani and Maurizio Magarini and Emil Bj{\"o}rnson},
	title   = {Parametric Channel Estimation With Short Pilots in {RIS}-Assisted Near- and Far-Field Communications},
	journal = {IEEE Trans. Wireless Commun.},
	year    = {2024},
	volume  = {23},
	number  = {8},
	pages   = {10366--10382},
	doi     = {10.1109/TWC.2024.3371715},
	month = {Aug.}
}

@article{jeongTVT2023jointCFOCE,
	author  = {Sumin Jeong and Arman Farhang and Nemanja Stefan Perovi{\'c} and Mark F. Flanagan},
	title   = {Joint {CFO} and Channel Estimation for {RIS}-Aided Multi-User Massive {MIMO} Systems},
	journal = {IEEE Trans. Veh. Technol.},
	year    = {2023},
	volume  = {72},
	number  = {9},
	pages   = {11800--11813},
	doi     = {10.1109/TVT.2023.3268247},
	month = {Sep.}
}

@misc{malekiArxiv2026limitedFeedback,
	author        = {Hossein Maleki and Hamid Jafarkhani},
	title         = {Channel-Estimation-Free Beamforming for {RIS}-Assisted Systems With Limited Feedback},
	year          = {2026},
	eprint        = {2608.01681},
	archivePrefix = {arXiv},
	primaryClass  = {eess.SP},
	note          = {arXiv:2608.01681},
	url           = {https://arxiv.org/abs/2608.01681},
	month = {Aug.}
}

@article{binWCL2026adaptiveCrossEntropy,
	author  = {Kai Bin and Yonggang Zhu and Yifu Sun and Kang An and Zhi Lin and Dusit Niyato},
	title   = {Adaptive Learning-Enhanced Cross-Entropy Algorithm for {RIS}-Aided Communications: Blind Beamforming Design and Experimental Validation},
	journal = {IEEE Wireless Commun. Lett.},
	year    = {2026},
	volume  = {15},
	pages   = {2283--2287},
	doi     = {10.1109/LWC.2026.3669175},
	month = {Mar.}
}

@article{chenWCL2024crossEntropy,
	author  = {Tianrui Chen and Minglei You and Yangyishi Zhang and Gan Zheng and Jean-Baptiste Gros and Geoffroy Lerosey and Youssef Nasser and Fraser Burton and Gabriele Gradoni},
	title   = {Model-Free Optimization and Experimental Validation of {RIS}-Assisted Wireless Communications Under Rich Multipath Fading},
	journal = {IEEE Wireless Commun. Lett.},
	year    = {2024},
	volume  = {13},
	number  = {3},
	pages   = {627--631},
	doi     = {10.1109/LWC.2023.3337709},
	month = {Mar.}
}

@article{yanTWC2025powerAutocorrelation,
	author  = {Ge Yan and Lipeng Zhu and Rui Zhang},
	title   = {Power Measurement Enabled Channel Autocorrelation Matrix Estimation for {IRS}-Assisted Wireless Communication},
	journal = {IEEE Trans. Wireless Commun.},
	year    = {2025},
	volume  = {24},
	number  = {3},
	pages   = {1832--1848},
	doi     = {10.1109/TWC.2024.3512668},
	month = {Mar.}
}

@article{riceTCOM2001frequencyEstimation,
	author  = {F. Rice and B. Cowley and B. Moran and M. Rice},
	title   = {{Cram{\'e}r--Rao} Lower Bounds for {QAM} Phase and Frequency Estimation},
	journal = {IEEE Trans. Commun.},
	year    = {2001},
	volume  = {49},
	number  = {9},
	pages   = {1582--1591},
	doi     = {10.1109/26.950345},
	month = {Sep.}
}

@article{pekcanOJCOMS2024elementwiseRIS,
	author  = {Dogan Kutay Pekcan and Ender Ayanoglu},
	title   = {Achieving Optimum Received Power for Discrete-Phase {RISs} With Elementwise Updates in the Least Number of Steps},
	journal = {IEEE Open J. Commun. Soc.},
	year    = {2024},
	volume  = {5},
	pages   = {2706--2722},
	doi     = {10.1109/OJCOMS.2024.3386559},
	month = {Apr.}
}

@misc{etsiWorkProgramme2025_2026,
	author = {{ETSI}},
	title  = {{ETSI} Work Programme 2025--2026},
	note   = {{p. 9. Accessed: Oct. 18, 2026}},
	url    = {https://www.etsi.org/e-brochure/work-programme/2025-2026/files/basic-html/page9.html}
}

@IEEEtranBSTCTL{IEEEfullAuthorNames,
	CTLdash_repeated_names = {no},
	CTLuse_forced_etal = {yes},
	CTLmax_names_forced_etal = {6},
	CTLnames_show_etal = {1}
}
\end{document}